\documentclass{PoS}

\usepackage{subfigure}
\usepackage{scalerel,stackengine}
\usepackage{graphicx}
\usepackage{sidecap}
\usepackage{wrapfig}
\usepackage{latexsym}
\usepackage{epsfig}
\usepackage[mathscr]{eucal}
\usepackage{amsfonts}
\usepackage{amscd}
\usepackage{cite}
\usepackage{array}
\usepackage{amssymb}
\usepackage{colordvi}
\usepackage[centertags]{amsmath}
\usepackage{enumerate}
\usepackage{graphicx}
\usepackage{booktabs}
\usepackage{theorem}
\usepackage[footnotesize]{caption}
\usepackage{soul}
\usepackage[makeroom]{cancel}
\usepackage{url}
\usepackage{slashed}
\usepackage{braket}
\usepackage{slashed}
\usepackage{color}
\usepackage{bbm}
\usepackage[utf8]{inputenc}
\newcommand{\newc}{\newcommand}
\newc{\be}{\begin{equation}}
\newc{\ee}{\end{equation}}
\newc{\bea}{\begin{eqnarray}}
\newc{\eea}{\end{eqnarray}}
\newc{\ol}{\overline}
\newc{\wt}{\widetilde}
\newc{\bs}{\boldsymbol}
\newc{\m}{\mathcal}
\newc{\la}{\langle}
\newc{\ra}{\rangle}
\usepackage{cleveref}
\crefname{chapter}{Chapter}{Chapter}
\crefname{section}{Sec.}{Secs.}
\crefname{table}{Tab.}{Tabs.}
\crefname{figure}{Fig.}{Figs.}
\crefname{equation}{Eq.}{Eqs.}
\crefname{appendix}{Appendix\ }{Appendix\ }

\newcommand{\beq}{\begin{eqnarray}}
\newcommand{\eeq}{\end{eqnarray}}
\newcommand{\bpmatrix}{\begin{pmatrix}}
\newcommand{\epmatrix}{\end{pmatrix}}
\newcommand{\ba}{\begin{array}}
\newcommand{\ea}{\end{array}}

\newenvironment{kasten*}[1]
{
\hspace{0.05\linewidth}
\begin{minipage}{0.95\linewidth}
\setlength{\fboxsep}{10pt}
\definecolor{shadecolor}{gray}{0.9}
\definecolor{framecolor}{gray}{0}

\MakeFramed {\FrameRestore}
\subsection*{#1}
}
{
\endMakeFramed
\end{minipage}
\vspace{1em}
}

\renewcommand{\ol}{\text{1l}}

\renewcommand{\Re}{\text{Re}\!}

\renewcommand{\eqref}[1]{Eq.~(\ref{#1})}

\newcommand{\bc}{\begin{center}}
\newcommand{\ec}{\end{center}}

\newcommand{\hour}{{~\text{h}}}

\newcommand{\nZ}{\mathbb{Z}}

\newcommand{\ii}{\mathit{i}}

\newcommand{\cbrak}[1]{\left(#1\right)}
\newcommand{\sbrak}[1]{\left[#1\right]}

\newcommand{\X}{\chi}
\newcommand{\mX}{m_{\chi}}

\newcommand{\gX}{g_{\chi}}
\newcommand{\GX}{G^{\chi}}
\newcommand{\mh}{m_{h_1}}
\newcommand{\mH}{m_{h_2}}

\allowdisplaybreaks

\usepackage[printonlyused]{acronym}

\usepackage{cleveref}
\crefname{chapter}{Chapter}{Chapter}
\crefname{section}{Sec.}{Secs.}
\crefname{table}{Tab.}{Tabs.}
\crefname{figure}{Fig.}{Figs.}
\crefname{equation}{Eq.}{Eqs.}
\crefname{appendix}{Appendix\ }{Appendix\ }

\title{NLO corrections to Vector Dark Matter Direct Detection - An update}

\ShortTitle{NLO corrections to VDM Direct Detection}

\author{Seraina Glaus\\
        Institute for Theoretical Physics, Karlsruhe Institute of Technology,  76128 Karlsruhe, Germany\\
         Institute for Nuclear Physics, Karlsruhe Institute of Technology, 76344 Karlsruhe, Germany
        E-mail: \email{seraina.glaus@kit.edu}\color{black}}

\author{Margarete M\"{u}hlleitner\\
        Institute for Theoretical Physics, Karlsruhe Institute of Technology, 76128 Karlsruhe, Germany\\
        E-mail: \email{milada.muehlleitner@kit.edu}\color{black}}

\author{Jonas M\"{u}ller\\
        Institute for Theoretical Physics, Karlsruhe Institute of Technology, 76128 Karlsruhe, Germany\\
        E-mail: \email{jonas.mueller@kit.edu}\color{black}}

\author{Shruti Patel\\
        Institute for Theoretical Physics, Karlsruhe Institute of Technology,  76128 Karlsruhe, Germany\\
         Institute for Nuclear Physics, Karlsruhe Institute of Technology, 76344 Karlsruhe, Germany
        E-mail: \email{shruti.patel@kit.edu}\color{black}}

\author{\speaker{Rui Santos}\thanks{I would like to thank the organisers for their partial support.}\\
   Centro de F\'{\i}sica Te\'{o}rica e Computacional,
   Faculdade de Ci\^{e}ncias,\\
   Universidade de Lisboa, Campo Grande, Edif\'{\i}cio C8
  1749-016 Lisboa, Portugal, and  \\
  ISEL - Instituto Superior de Engenharia de Lisboa,\\
  Instituto Polit\'ecnico de Lisboa
 1959-007 Lisboa, Portugal.\\
        E-mail: \email{rasantos@fc.ul.pt}\color{black}}

\usepackage{xcolor}
\usepackage{changes}
\abstract{
In this work we present an update to a previous calculation 
of the Next-to-Leading Order (NLO) corrections to the Vector Dark Matter (VDM) direct detection cross section.
The model under investigation is a minimal extension of the Standard Model (SM) with one extra vector boson
and  one extra complex scalar field, where the vector is the DM candidate. We have computed the spin-independent cross section for the scattering of the VDM particle with
a nucleon. We now provide an update to the NLO cross section for the direct detection of the DM particle. 
We  further discuss the phenomenological implications of the NLO
corrections for the sensitivity of the direct detection DM 
experiments. }

\FullConference{Corfu Summer Institute 2019 "School and Workshops on Elementary Particle Physics and Gravity" (CORFU2019)\\
		31 August - 25 September 2019\\
		Corfù, Greece}

\begin{document}

\section{Introduction\label{sec:Intro}}
The experimental evidence for the existence of Dark Matter (DM) can be traced back to the work of Zwicky~\cite{Zwicky:1933gu}, "The redshift of extragalactic nebulae".
Many experimental results from different sources and origins have been
accumulated over the years leading to the conclusion that 27\% of the energy density of the Universe 
is unaccounted for in the Standard Model (SM) of particle physics, and this missing dark matter
is most likely a particle. These results are all gravitational in origin
which means that the properties of dark matter are dictated by Astronomy and Cosmology.
If DM is indeed a particle it could be produced at colliders, but signatures of missing
energy alone cannot be a proof of the existence of DM. Indirect searches for DM annihilation
can also hint at the existence of DM and in the worst case scenario they can at least be used to
exclude specific models or regions of their parameter space.
 Direct DM detection
is our best hope to unambiguously find a dark matter candidate and it is therefore the place where precision matters the most. When searching for DM in direct detection,
all available experimental data combined favours a weakly  interacting massive particle (WIMP) with a velocity of the order of 200 km/s. 
 In this work we discuss a minimal model which is
 an extension of the SM by the addition of a dark vector  $\chi_\mu$ with a gauged
$U(1)_{\chi}$ symmetry and a complex SM-gauge singlet
$S$. We will call this model Vector Dark Matter (VDM) 
in the following.

As shown in~\cite{Goodman:1984dc}, DM particles that undergo coherent scattering with 
nuclei are the easiest to detect due the larger scattering rates.  There are many uncertainties from
cosmological and astronomical origins but particle physicists have tried to increase precision from their
side by calculating higher-order corrections to the scattering cross sections, both
strong and electroweak~\cite{Haisch:2013uaa, Crivellin:2014qxa, Hill:2014yka,
  Abe:2015rja, Klasen:2016qyz, Azevedo:2018exj, Ishiwata:2018sdi,
  Ghorbani:2018pjh, Abe:2018emu, Ertas:2019dew, Glaus:2019itb}. 
 %
The electroweak corrections to the coherent scattering of the DM
candidate $\chi_\mu$ require the renormalisation of the VDM
model. After renormalisation, the coefficients from each term in the spin-independent 
amplitude, with renormalised loop corrections included, are matched 
to the effective couplings of the Lagrangian, ${\cal L}_{\rm eff}$, which describes the coupling of two DM
particles with two quarks. These will be the corrected coefficients 
to the corresponding tree-level effective couplings from ${\cal L}_{\rm eff}$. 

Most of the work presented here was published in Ref.~\cite{Glaus:2019itb}. We have extended the work by calculating one-loop corrections
to the $\bar q q h$ vertex. This will be discussed in detail in \cref{sec:lvcorrections}.
The main conclusions are the same as in our previous work~\cite{Glaus:2019itb}, except for a slight reduction in the overall NLO corrections
relative to the LO result.



\section{The Vector Dark Matter Model\label{sec:VDM}}

In this section we briefly review our VDM model and refer the reader to Refs.~\cite{Glaus:2019itb, Hambye:2008bq, Lebedev:2011iq, Farzan:2012hh, Baek:2012se,
Baek:2014jga, Duch:2015jta, Azevedo:2018oxv, YaserAyazi:2019caf} for
details. 
The model has two new fields relative to the SM: one vector boson and one complex scalar singlet.
Besides the SM symmetries there is now a new $U(1)_{\X}$
 gauge symmetry under which all SM fields are neutral. The new singlet is a scalar under the SM gauge group but has unit charge under  $U(1)_{\X}$.
The appearance of the new dark gauge boson, named $\chi_\mu$,  is a consequence of the gauged $U(1)_{\X}$ symmetry.
In order to have a stable VDM candidate we further force the
 model to be invariant 
under the  $\nZ_2$ symmetry,
\begin{equation}
	X_\mu \to - X_\mu, \quad \mathbb{S} \to \mathbb{S}^*
\end{equation}
 for the dark gauge boson $\chi_{\mu}$ and for the
 singlet field $\mathbb{S} $ . The SM particles are all even 
under $\nZ_2$, and therefore there is no kinetic mixing between the gauge bosons from $U(1)_{\X}$ and from the
SM $U(1)_Y$. The complete Lagrangian of the theory is
\begin{equation}
{\cal L} =  {\cal L}_{SM} -  \frac{1}{4} X_{\mu \nu} X^{\mu \nu} + (D_\mu \mathbb{S})^\dagger (D^\mu \mathbb{S}) + \mu_S^2 \left| \mathbb{S} \right|^2 - \lambda_S \left| \mathbb{S} \right|^4 
- \kappa  \left| \mathbb{S} \right|^2 H^\dagger H \,,
\end{equation}
where $X^{\mu \nu}$ is the $U(1)_{\X}$ field-strength tensor and the covariant derivative
\begin{equation}
D_{\mu}  \mathbb{S} = \left(\partial_{\mu}+  \mathit{i} \gX \chi_{\mu}\right)  \mathbb{S} \,,
\end{equation}
with $\gX$ being the gauge coupling of the dark gauge boson
$\chi_{\mu}$. 
The mass and coupling parameters $\mu_S^2$,
  $\lambda_S$ and $\kappa$ are all real.
The SM potential has the form $ V_{SM}= -\mu_H^2 \vert H\vert^2 +\lambda_H \vert
H\vert^4$.
Both the neutral component of the doublet $H$ and the real part of
the singlet field $\mathbb{S} $ acquire vacuum
expectation values (VEV) $v$ and 
$v_S$, respectively. They are expanded around their VEVs as
\begin{equation}
	H = 
	\begin{pmatrix}
	G^{+}\\\frac{1}{\sqrt 2}\cbrak{v + \Phi_H +\mathit{i} \sigma_H}
	\end{pmatrix} \quad \mbox{and} \quad
	 \mathbb{S} = \frac{1}{\sqrt{2}} \left( v_S + \Phi_S+ \mathit{i} \sigma_S\right)\,,
\end{equation}
where $\Phi_H$ and $\Phi_S$ 
  denote the CP-even field components of $H$ and
 $\mathbb{S}$, respectively.

The imaginary components of the doublet, $\sigma_H$, and of the singlet, $\sigma_S$,
are the neutral SM-like Goldstone boson $G^0$ and the Goldstone boson $\GX$ for the
gauge boson $\chi_{\mu}$, respectively. The charged Goldstone boson, partner of the 
$W^\pm$ boson, is $G^{\pm}$.  We write the minimum conditions as
\begin{align}
	\left< \frac{\partial V}{\partial \Phi_H}\right>\equiv
  \frac{T_{\Phi_H}}{v} &= \left(\frac{\kappa v_S^2 }{2}+\lambda_H
                         v^2-\mu_H^2\right)\,, \\
    \left<\frac{\partial V}{\partial \Phi_S}\right>\equiv
  \frac{T_{\Phi_S}}{v_S} &= \left(\frac{\kappa v^2 }{2}+\lambda_S
                           v_S^2-\mu_S^2\right)\,, 
    \label{VDM::tadpoles}
\end{align}
which in turn allows us to express the mass matrix of the scalar particles as 
\begin{equation}
	\mathcal{M}_{\Phi_h \Phi_S} =
	\begin{pmatrix}
		2 \lambda_H v^2 & \kappa v v_S\\
		\kappa v v_S & 2 \lambda_S v_S^2
	\end{pmatrix}
	+
	\begin{pmatrix}
		\frac{T_{\Phi_H}}{v} & 0\\
		0 & \frac{T_{\Phi_S}}{v_S}
    \end{pmatrix}\,.
    \label{VDM::massmatrix}
\end{equation}
The CP-even mass eigenstates $h_1$ and $h_2$ are then obtained via the rotation matrix
$R_\alpha$ as
\begin{equation}
    \begin{pmatrix}
        h_1 \\ h_2
    \end{pmatrix}
    =
    R_{\alpha} \begin{pmatrix}
        \Phi_H \\ \Phi_S
    \end{pmatrix}
    \equiv
    \begin{pmatrix}
        \cos\alpha & \sin\alpha \\
        -\sin\alpha & \cos\alpha
    \end{pmatrix}
    \begin{pmatrix}
        \Phi_H \\ \Phi_S
    \end{pmatrix} \;.
    \label{VDM::massbasis} 
\end{equation}
The physical scalar states are $h_1$ and $h_2$ with masses $m_{h_1}$ and $m_{h_2}$. Denoting 
the mass of the VDM particle by $m_{\chi}$ we choose the following set of independent parameters  
\begin{equation}
    m_{h_1}\,,m_{h_2}\,,\mX\,,\alpha\,,v\,,\gX\,, T_{\Phi_H}\,, T_{\Phi_S}\,.
\end{equation}
The remaining parameters can be written as a function of this set as 
\beq
\label{VDM::treelevelrelations1}
\lambda_H &=& \frac{m_{h_1}^2\cos^2\alpha+m_{h_2}^2 \sin^2\alpha}{2 v^2}\,, \\
\kappa &=& \frac{\left(m_{h_1}^2-m_{h_2}^2\right)\cos\alpha\sin\alpha}{v
  v_S}\,, \label{VDM::treelevelrelations2}\\
\lambda_S &=& \frac{m_{h_1}^2\sin^2\alpha+m_{h_2}^2 \cos^2\alpha}{2 v_S}\,,
\label{VDM::treelevelrelations3}\\
v_S &=&\frac{\mX}{\gX}  \label{VDM::treelevelrelations4} \,.
\eeq
The SM VEV $v \approx 246$~GeV is fixed by the $W$ boson mass and the mixing angle $\alpha$ is varied in the interval
$-\frac{\pi}{2} \leq \alpha < \frac{\pi}{2}$. We require the potential to be in a global minimum, that perturbative unitarity holds
and enforce the potential to be bounded from below implying the conditions,
\beq
\lambda_H > 0, \ \ \lambda_S >0, \ \ \kappa > -2 \sqrt{\lambda_H \lambda_S}.
\eeq

\subsection{Renormalisation of the VDM Model \label{sec:Renorm}}
In this section we briefly highlight the renormalisation procedure and direct the reader to Ref.~\cite{Glaus:2019itb}
for details. There are four new parameters relative to the SM that need to be renormalised: the non-SM-like scalar 
mass, $m_{h_2}$, the rotation angle $\alpha$, the coupling  $\gX$ and the DM mass
$\mX$.\footnote{Note that in our notation $h_1$ corresponds to the
    SM-like Higgs boson, while we attribute $h_2$ to the non-SM-like scalar.} Our renormalisation
 procedure is the following. Once the free parameters are chosen we replace 
the bare parameters $p_0$ with the renormalised
ones $p$ according to  
\begin{equation}
    p_0=p + \delta p\,,
\end{equation}
where $\delta p$ is the counterterm for the parameter $p$. The fields $\Psi$ are renormalised multiplicatively,
\begin{equation}
    \Psi_0 = \sqrt{Z_{\Psi}} \Psi\,,
    \label{renorm::Zfactorgen}
\end{equation}
where $Z_\Psi$ is the field renormalisation constant and  $\Psi_0$ stands for the bare field and $\Psi$ for the renormalised
field. When there is mixing like is the case for our scalar sector, $\sqrt{Z_{\Psi}}$ is a matrix. 

The renormalisation of the SM is by now a textbook subject. Therefore we will discuss only the renormalisation
of the extra parameters of the model. In the gauge sector there is just one extra field, that is, one extra mass
renormalisation constant and one field renormalisation constant. 
Furthermore, the $\nZ_2$ symmetry under which only the dark gauge boson
$\chi_{\mu}$ is odd, precludes kinetic mixing between the gauge bosons of the 
$U(1)_{\X}$ and that of the $U(1)_Y$. Since the symmetry is broken only spontaneously
this is true to all orders in perturbation theory. 
We define $\mX^2 \rightarrow \mX^2 + \delta \mX^2$
and $ \chi  \rightarrow \cbrak{1+\frac{1}{2}\delta Z_{\chi\chi}} \chi$ and the on-shell (OS) conditions yield the following expressions
for the counterterms 
\begin{equation}
\delta Z_{\chi\chi} =
    -\Re\, \frac{\partial\Sigma_{\chi\chi}^T(p^2)}{\partial
      p^2}\bigg\vert_{p^2=m_\chi^2} \,\quad\text{,}\quad     \quad\text{and}\quad
    \delta \mX^2 = \Re~\Sigma_{\chi\chi}^{T}\cbrak{\mX^2}\,, 
\end{equation}
where the subscript $T$ identifies the transverse part of the self-energies

The dark gauge coupling $\gX$ cannot be measured directly in a physical process and we have
therefore decided to renormalise it using the $\overline{\mbox{MS}}$ scheme. We choose
the triple vertex  $h_1h_1h_1$ to determine $\gX$ (the UV divergence is universal).
Defining
\begin{equation}
    \mathcal A^{\text{NLO}}_{h_1h_1h_1} = \mathcal A^{\text{LO}}_{h_1h_1h_1} + \mathcal A^{\text{VC}}_{h_1h_1h_1} +\mathcal A^{\text{CT}}_{h_1h_1h_1}\,,
\end{equation}
where $\mathcal{A}^{\text{VC}}$ and $\mathcal A^{\text{CT}}$ are the amplitude for the
virtual corrections and vertex counterterms, respectively. Dropping the index  $h_1h_1h_1$, the counterterm amplitude can be written as
\begin{equation}
    \mathcal{A}^{\text{CT}} = \delta^{\text{mix}} + \delta g^{\text{CT}}
\end{equation}
with 
\begin{equation}
    \delta^{\text{mix}} = \frac{3}{2} g_{h_1h_1h_1} \delta Z_{h_1h_1} + \frac{3}{2} g_{h_1h_1h_2} \delta Z_{h_2h_1}    
\end{equation}

The trilinear Higgs self-coupling reads
\beq
g_{h_1 h_1 h_1 } = - \frac{3 g m_{h_1}^2}{2 m_W} \cos^3\alpha - \frac{3 
\gX m_{h_1}^2}{\mX} \sin^3\alpha \;.
\eeq
and the corresponding CTs are 
\begin{equation}
    \delta g^{\text{CT}} = \sum_p\frac{\partial
      g_{h_1h_1h_1}}{\partial p} \delta p\,,\quad p=
 \mh^2,m_\chi^2,m_W^2,g,\alpha,\gX\,. 
\end{equation}

The divergent part of $ \delta \gX$ is then given by 
\begin{equation}
    \delta \gX\big\vert_{\text{div}} = \cbrak{\frac{\mX}{3 \mh^2\sin^3\alpha}} \cbrak{\mathcal A^{\text{VC}}+\mathcal{A}^{\text{CT}}\big\vert_{\delta \gX =0}}\big\vert_{\text{div}} \,,
    \label{renorm::CTgX}
\end{equation}
where $\cbrak{\dots}_{\text{div.}}$ indicates the UV pole. 

The one-loop diagrams were generated with {\tt
  FeynArts}\cite{Hahn:2000kx} for which the model file was obtained with {\tt SARAH} 
\cite{Staub:2013tta,Staub:2012pb,Staub:2010jh,Staub:2009bi} and the
program packages {\tt FeynCalc}~\cite{Shtabovenko:2016sxi,Mertig:1990an} 
and  {\tt FormCalc}~\cite{Hahn:1998yk} 
were used to reduce the
amplitudes to Passarino-Veltmann integrals
\cite{Passarino:1978jh}. The numerical evaluation of the integrals was
done by {\tt Collier}
\cite{Denner:2016kdg,Denner:2002ii,Denner:2005nn,Denner:2010tr}. 
The counterterm $\gX$ in the $\overline{\mbox{MS}}$ scheme is then obtained as
\begin{equation}
    \delta \gX\big\vert_{\varepsilon} = \frac{\gX^3}{96\pi^2}\Delta_{\varepsilon}\,,
\end{equation}
with 
\beq
\Delta_{\varepsilon} = \frac{1}{\varepsilon} - \gamma_E +\ln
4\pi \;,
\eeq
where $\gamma_E$ denotes the Euler-Mascheroni constant. 

We end this section with the renormalisation of the scalar sector. Relative to the SM we have a new field, the real component $\Phi_S$ of the
singlet, which mixes with the real neutral $\Phi_H$ of the Higgs doublet. These two fields mix giving rise to two mass eigenstates
$h_1$ and $h_2$ and a mixing angle $\alpha$. Hence, the field renormalisation constants are now written as  
\begin{equation}
    \begin{pmatrix}
        h_1 \\ h_2
    \end{pmatrix}
    \rightarrow
    \begin{pmatrix}
        1+\frac{1}{2}\delta Z_{h_1 h_1} & \frac{1}{2}\delta Z_{h_1 h_2}\\
        \frac{1}{2} \delta Z_{h_2 h_1} & 1+\frac{1}{2} \delta Z_{h_2 h_2}
    \end{pmatrix}
    \begin{pmatrix}
        h_1 \\ h_2
    \end{pmatrix}\,.
    \label{renorm::ZfactorHiggs}
\end{equation}
In the mass eigenbasis, the mass matrix in \cref{VDM::massmatrix} yields
\begin{equation}
    \mathcal{M}_{h_1 h_2} = 
    \underbrace{
    \begin{pmatrix}
        m_{h_1}^2 & 0\\
        0 & m_{h_2}^2
    \end{pmatrix}
    }_{\equiv \mathcal{M}^2}
    +
    \underbrace{
    R_{\alpha} \begin{pmatrix}
        T_{\Phi_H}/v & 0 \\
        0& T_{\Phi_S}/v_S
    \end{pmatrix}
    R_{\alpha}^T
    }_{\equiv\delta T}\,.
    \label{RENORM::MASSMATRIX}
\end{equation}
The tadpole terms in the tree-level mass matrix are \textit{bare
  parameters}. Therefore we first have to renormalise the tadpoles in
  such a way that the theory has a minimum at next-to-leading order
  (NLO).
The tadpole renormalisation condition counterterms is defined as
\begin{equation}
    \hat T_i =T_i - \delta T_i \overset{!}{=}0\,,\quad
    i =  \Phi_H , \Phi_S \,,
\end{equation}
where  $\hat T_i$ is the one-loop renormalised tadpole.
In the mass basis the tadpole counterterms are written as 
\begin{equation}
    \begin{pmatrix}
        T_{h_1}\\T_{h_2}
    \end{pmatrix}
    =
    R_{\alpha}\cdot 
    \begin{pmatrix}
    T_{\Phi_h}\\T_{\Phi_S}
    \end{pmatrix}\,.
    \label{RENORM::TadpolCondi}
\end{equation}
which in turn implies
\begin{equation}
    \delta \mathcal{M}_{h_1 h_2} =
    \begin{pmatrix}
    \delta m_{h_1}^2 & 0 \\
    0&\delta m_{h_2}^2    
    \end{pmatrix}
    + R_{\alpha} 
    \begin{pmatrix}
        \frac{\delta T_{\Phi_H}}{v} & 0 \\
        0 & \frac{\delta T_{\Phi_S}}{v_S}
    \end{pmatrix}
    R_{\alpha}^T
    \equiv 
    \begin{pmatrix}
        \delta m_{h_1}^2 & 0 \\
        0&\delta m_{h_2}^2 
        \end{pmatrix}
    +  \begin{pmatrix}
        \delta T_{h_1 h_1} & \delta T_{h_1 h_2} \\
        \delta T_{h_2 h_1}&\delta T_{h_2 h_2}    
        \end{pmatrix}\,,
        \label{RENORM::MASSCOUNTERTERM}
 \end{equation}
In \cref{RENORM::MASSCOUNTERTERM} we neglect all terms of order
$\mathcal O \cbrak{\delta\alpha\delta T_i}$ since they
are formally of two-loop order. Using OS conditions and
\cref{RENORM::MASSCOUNTERTERM} finally yields the
following relations for the counterterms ($i=1,2$) 
\begin{align}
    &\delta m^2_{h_i} = \Re\sbrak{\Sigma_{h_ih_i}(m_{h_i}^2) - \delta T_{h_ih_i}}\,,\\
    &\delta Z_{h_ih_i}  = -
      \Re\sbrak{\frac{\partial\Sigma_{h_ih_i}(p^2)}{\partial
      p^2}}_{p^2=m_{h_i}^2}\,,\\ 
    &\delta Z_{h_ih_j} = \frac{2}{m_{h_i}^2-m_{h_j}^2}
      \Re\sbrak{\Sigma_{h_ih_j}(m_{h_j}^2)-\delta T_{h_ih_j}}\,,\quad
      i\neq j\,.\label{RENORM::ZFACTOR} 
\end{align}

We now move to the final parameter that needs to be renormalised, the mixing angle
$\alpha$. There are processes that depend on the mixing angle and so one option
is to use one such process 
This leads, however, to
unphysically large
counterterms~\cite{Krause:2016oke}. The renormalisation 
of the mixing angles in SM extensions was thoroughly discussed
in~\cite{Krause:2016oke, Bojarski:2015kra, Denner:2016etu,Krause:2016xku,Krause:2017mal,Altenkamp:2017ldc, Altenkamp:2017kxk, Fox:2017hbw,
  Denner:2018opp, Grimus:2018rte,Krause:2018wmo,Krause:2019oar}. In this
work we will use the scheme proposed in~\cite{Pilaftsis:1997dr, Kanemura:2004mg}, which
connects the derivation of the angle counterterm with the usual OS conditions of the 
scalar field to the relations between the gauge basis and the mass
basis. The bare parameter expressed through the renormalised one and
the counterterm reads
\begin{equation}
    \alpha_0 = \alpha + \delta \alpha\,.
\end{equation}
Considering the field strength renormalisation before the  rotation, 
\begin{equation}
    \begin{pmatrix}
        h_1\\h_2
    \end{pmatrix}
    = R\cbrak{\alpha+\delta\alpha} \sqrt{Z_{\Phi}}\begin{pmatrix}
        \Phi_H\\\Phi_S
    \end{pmatrix}\,,
\end{equation}
and expanding it to strict one-loop order, 
\begin{equation}
    R\cbrak{\alpha+\delta\alpha} \sqrt{Z_{\Phi}}
    \begin{pmatrix}
        \Phi_H\\\Phi_S
    \end{pmatrix}
    =\underbrace{R(\delta\alpha) R(\alpha)
      \sqrt{Z_{\Phi}}R(\alpha)^T}_{\overset{!}{=}\sqrt{Z_H}} R(\alpha)  
    \begin{pmatrix}
        \Phi_H\\\Phi_S
    \end{pmatrix}
    +\mathcal{O} (\delta\alpha^2)
    = \sqrt{Z_H}   
    \begin{pmatrix}
        h_1\\h_2
    \end{pmatrix}\,,
\end{equation}
yields the field strength renormalisation matrix $\sqrt{Z_H} $
connecting the bare and renormalised fields in the mass basis. 
This finally leads to the condition~\cite{Krause:2016oke}
\begin{eqnarray}
    \delta \alpha & = & \frac{1}{4}\cbrak{\delta Z_{h_1h_2}-\delta Z_{h_2h_1}} \\
    &&=
       \frac{1}{2(m_{h_1}^2-m_{h_2}^2)}\Re\cbrak{\Sigma_{h_1h_2}(m_{h_1}^2)+
     \Sigma_{h_1h_2}(m_{h_2}^2)- 2 \delta T_{h_1h_2}}\,. 
\end{eqnarray}

In the numerical analysis presented in our work~\cite{Glaus:2019itb} we have used two 
further renormalisation schemes for $\delta\alpha$: the $\overline{\mbox{MS}}$ scheme and a
process-dependent scheme. The results presented here use, however, only the scheme
previously described.

\section{Dark Matter Direct Detection at Tree Level\label{sec:ddtree}}

The spin-independent (SI) cross section of DM-nucleon scattering
can be described with an effective Lagrangian.  The largest
contributions  to the cross section are due  to light quarks
$q=u,d,s$ and gluons. In the VDM model the SI cross section is well described by  
an effective operator Lagrangian \cite{Hisano:2010yh} 
\begin{equation}
    \mathcal L ^{\text{eff}} = \sum_{q=u,d,s}\mathcal L^{\text{eff}}_q + \mathcal
    L^{\text{eff}}_G \;,
    \label{HISANO::LEFF}
\end{equation}
with 
\begin{subequations}    
    \begin{align}
        &\mathcal L ^{\text{eff}}_q = f_q \chi_{\mu}\chi^{\mu} m_q
          \bar q q  +
          \frac{g_q}{\mX^2}\chi^{\rho}\ii \partial^{\mu}\ii\partial^{\nu}\chi_{\rho}
         \mathcal{O}^q_{\mu\nu}\,,\\  
        &\mathcal L^{\text{eff}}_G = f_G \chi_{\rho}\chi^{\rho}
          G^{a}_{\mu\nu}G^{a \,\mu\nu}\,, \label{eq:ggxx}
    \end{align}
    \label{tree::effoperators}
\end{subequations}
where $G_{\mu\nu}^a$ ($a=1,...,8$) is the gluon field strength tensor and $\mathcal{O}_{\mu\nu}^q$ is the quark twist-2
operator corresponding to the traceless part of the energy-momentum
tensor of the nucleon \cite{Hisano:2010ct,Hisano:2015bma}, 
\begin{equation}
    \mathcal{O}_{\mu\nu}^q = \frac{1}{2}\bar q \ii
    \cbrak{\partial_{\mu}\gamma_{\nu}+\partial_{\nu}\gamma_{\mu}-
    \frac{1}{2}\slashed{\partial}}q\,.  
\end{equation}
In our calculation we will neglect operators suppressed by the DM velocities and
also the gluon twist-2 operator $\mathcal
O_{\mu\nu}^g$, because they are one order higher in the
strong coupling constant $\alpha_s$ \cite{Hisano:2010yh}. 

Taking the nucleon states to be on-shell and considering vanishing momentum transfer, the
nucleon matrix elements are given by 
\begin{subequations}    
\beq
        \bra{N}m_q\bar q q \ket{N} &=& m_N f^N_{T_q} \\
       - \frac{9\alpha_S}{8\pi} \bra{N}G_{\mu\nu}^a
        G^{a,\mu\nu}\ket{N} &=& \left(1 - \sum_{q=u,d,s}
          f^N_{T_q}\right) m_N = m_N f^N_{T_G} \\
        \bra{N(p)}\mathcal O_{\mu\nu}^q\ket{N(p)} &=&
        \frac{1}{m_N}\cbrak{p_{\mu}p_{\nu}-\frac{1}{4}m_N^2
          g_{\mu\nu}}\cbrak{q^N(2)+\bar q^N(2)} \;,
\eeq
    \label{HISANO::MATRIXELEMENT}
\end{subequations}
where $N$ stands for a nucleon, $N=p,n$, and $m_N$ is the 
nucleon mass and $p$ is the four-momentum  of the nucleon. The fraction of momentum carried by the quarks 
is determined by  the second moments, $q^N(2)$ and $\bar q^N(2)$, of the parton distribution functions of the quark $q(x)$ and
the antiquark $\bar q(x)$, respectively. $f^N_{T_q}$, $f^N_{T_G}$ denote the fraction of the nucleon mass that is due to 
light quarks $q$ or to the gluon, respectively. These are obtained from lattice calculations and are given  in App.~\ref{APP::NUCLEAR}\color{black}. 

\begin{figure}
    \centering
    \includegraphics[width=0.2\textwidth]{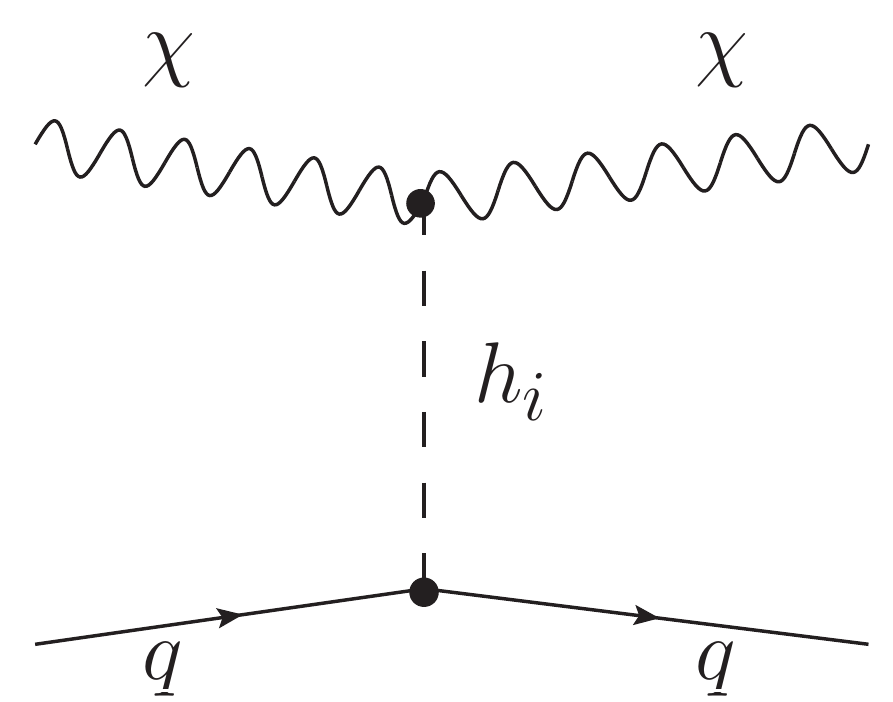}
\vspace*{0.2cm}
    \caption{Generic tree-level diagram contribution to the SI cross
      section. The mediator $h_i$ corresponds to the two Higgs bosons
      $h_1$ and $h_2$. The quark line $q$ corresponds to all quarks
      $q=u,d,s,c,b,t$. \label{fig:trelevdiag}} 
\end{figure}

The SI scattering DM-nucleon cross section can now be written as 
\begin{equation}
    \sigma_N = \frac{1}{\pi}\cbrak{\frac{m_N}{\mX+m_N}}^2\big\vert f_N\big\vert^2\,,
    \label{TREE::CX}
\end{equation}
where the nucleon is either a proton or a neutron ($N=p,n$) and 
\begin{equation}
    f_N/m_N = \sum_{q=u,d,s} f_q f^N_{T_q} + \sum_{q=u,d,s,c,b}
    \frac{3}{4}\cbrak{q^N(2)+\bar q^N(2)} g_q -\frac{8\pi}{9\alpha_S}
    f^N_{T_G}f_G\,.
\label{eq:fnovermn}
\end{equation}
In the contribution from the quark twist-2 operator all quarks
below the energy scale $\sim 1$~GeV have to be included,
{\it i.e.}~all quarks but the top quark.
\begin{figure}[h!]
    \centering
    \includegraphics[width=0.2\textwidth]{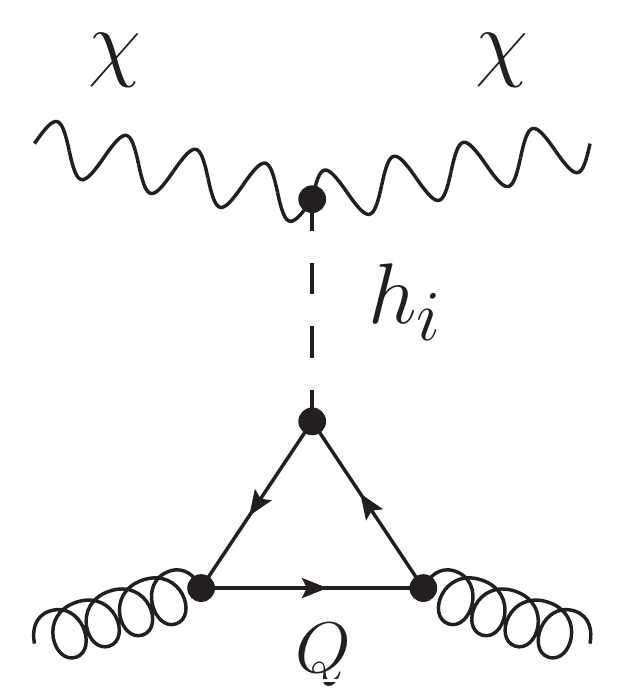}
\vspace*{0.2cm}
    \caption{Higgs bosons $h_i$ mediating the coupling of two
      gluons to two VDM particles through a heavy
      quark loop. \label{fig:gluontriangle}} 
\end{figure} 
The sum in the first term of Eq.~(\ref{eq:fnovermn}\color{black}) is only
over the light quarks. There is, however, a leading-order gluon interaction
through a heavy quark triangle diagram, {\it
  cf.}~Fig.~\ref{fig:gluontriangle}\color{black}, with a charm, bottom
  or top quark in the loop. Since their mass is above the energy scale 
  relevant for DM direct detection, they should be integrated out for the description
of the interaction at the level of the nucleon.  This is done by calculating the
heavy quark triangle diagrams and then integrating out the heavy
quarks. The procedure is equivalent to calculating the amplitude in Fig.~\ref{fig:trelevdiag} \color{black}
with heavy quarks $Q = c,b,t$, and replacing the resulting tensor
structure $m_Q  \bar{Q}Q$ with the effective gluon operator
\cite{Shifman:1978zn,Ertas:2019dew,Abe:2018emu} 
\begin{equation}
    m_Q \bar Q Q \rightarrow - \frac{\alpha_S}{12\pi} G_{\mu\nu}^aG^{a\mu\nu}\,,
    \label{tree::gluonmap}
\end{equation}
corresponding to the effective leading-order VDM-gluon interaction in
\cref{tree::effoperators}.

The tree-level diagrams contributing to the SI cross section are
shown in Fig.~\ref{fig:trelevdiag} \color{black} and are calculated for
vanishing momentum transfer. The Wilson coefficient for each effective operator
in \cref{HISANO::LEFF}  is extracted by projecting 
onto the corresponding tensor structure, $m_q q \bar{q}$, leading to
\begin{equation}
    f_q = \frac{1}{2} \frac{g
      \gX}{m_W}\frac{\sin(2\alpha)}{2}\frac{\mh^2-\mH^2}{\mh^2\mH^2}
    \mX\,,\quad q=u,d,s,c,b,t\,.
\label{eq:fqlo}
\end{equation}
As previously discussed, the heavy quarks $Q=b,c,t$ contribute to the effective gluon interaction and using
Eq.~(\ref{tree::gluonmap} \color{black}), the Wilson coefficient for the gluon interaction, $f_G$, can be written in 
terms of $f_q$ for $q=c,b,t$,
\beq
f_G = \sum_{q=c,b,t} - \frac{\alpha_S}{12\pi} f_q \;,
\eeq
resulting in the SI LO cross section
\begin{equation}
    \sigma^{\text{LO}} = \frac{\sin^22\alpha}{4\pi}\cbrak{\frac{\mX
        m_N}{\mX+m_N}}^2\frac{\cbrak{\mh^2-\mH^2}^2}{\mh^4\mH^4}\,\,\frac{\mX^2
      m_N^2}{v^2 v_S^2}\,\left| \sum_{q=u,d,s}
    f^N_{T_q}+3 \cdot \frac{2}{27}f^N_{T_G}\right|^2\,. \label{eq:siddlo}
\end{equation}
The twist-2 operator does not contribute at LO.

\section{Dark Matter Direct Detection at One-Loop
  Order\label{sec:dd1loop}} 

\begin{figure}
    \centering
    \subfigure[UpV
    Corrections\label{VERTEXCORRECTIONS}]{\includegraphics[width=0.23\textwidth]{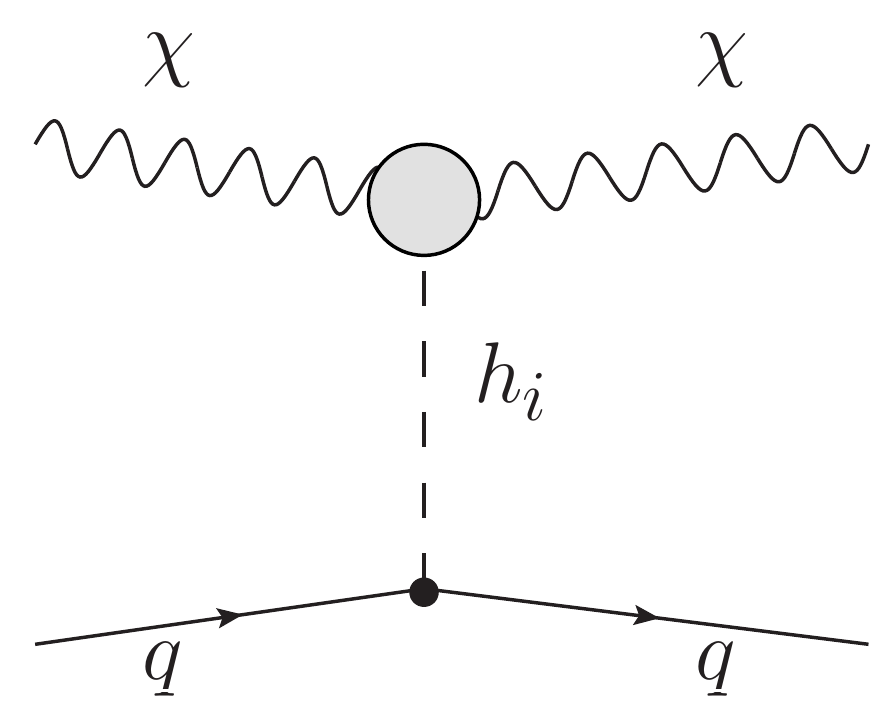}} \hspace*{0.4cm}
    \subfigure[Mediator
    Corrections\label{MEDIATORCORRECTIONS}]{\includegraphics[width=0.23\textwidth]{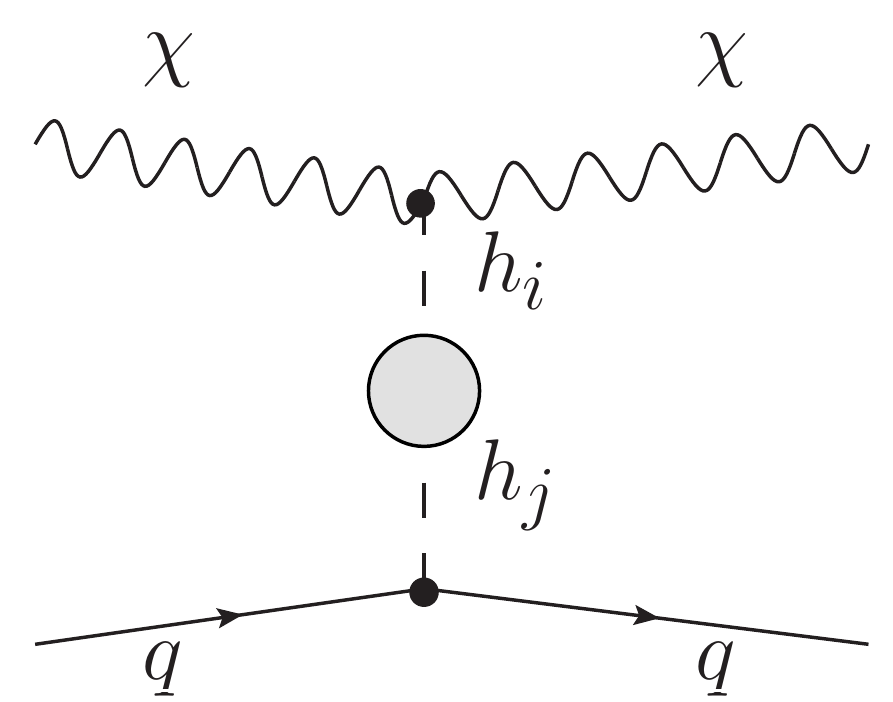}} \hspace*{0.4cm}
    \subfigure[Box Corrections\label{BOXCORRECTIONS}]{\includegraphics[width=0.17\textwidth]{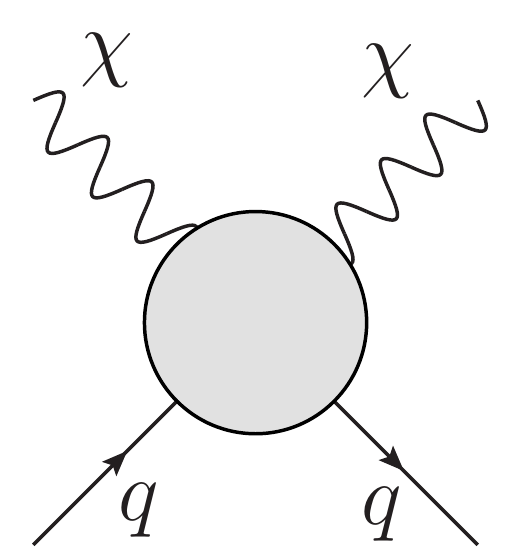}}\hspace*{0.4cm}
     \subfigure[LoV Corrections\label{LVCORRECTIONS}]{\includegraphics[width=0.23\textwidth]{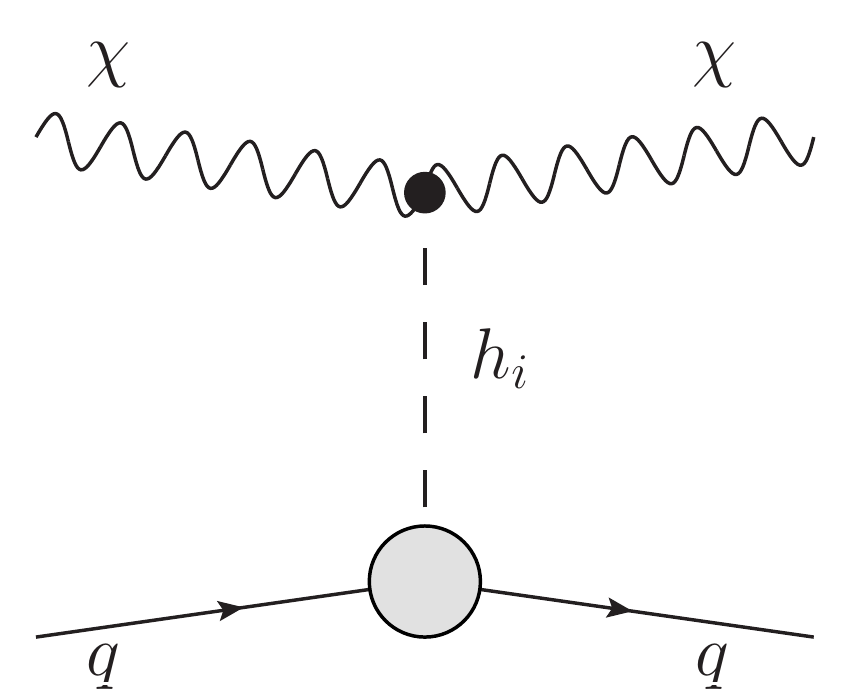}}
    \caption{Generic one-loop corrections to the scattering of VDM
      with the nucleon. The grey blob corresponds to the renormalized
      one-loop corrections. The corrections can be separated into
      upper vertex (a), mediator (b), box (c) and lower vertex (d) corrections.}
    \label{OneLoop::Tops}
\end{figure}

Let us now calculate the NLO~electroweak (EW) contribution to the cross section. Here again we will just briefly review 
our calculation in \cite{Glaus:2019itb} and present some updates to the calculation. We need to determine
the Wilson coefficients $f_q$ and $f_G$ related to the operators in \cref{tree::effoperators}.
At NLO~EW also $g_q$ contributes to the cross section.
The diagrams contributing at NLO~EW are shown in \cref{OneLoop::Tops}.  

In our study presented in \cite{Glaus:2019itb} we have not included the contributions of the diagrams in \cref{LVCORRECTIONS}.
These were now included and the results presented here are updated. The treatment of the diagrams will be discussed in detail in \cref{sec:lvcorrections}.



\subsection{Upper Vertex Corrections $\chi\chi h_i$}
The effective one-loop coupling $\chi\chi h_i$ is extracted 
from the loop corrections to the $\chi\chi
  h_i$  coupling. We take the DM particles on-shell and assume $p_{h_i}=0$.
The amplitude for the NLO vertex (in this section we will refer to the upper vertex as just the vertex) can be written as  
\begin{equation}
    \ii \mathcal A_{\chi\chi h_i}^{\text{NLO}} = \ii
    \mathcal{A}_{\chi\chi h_i}^{\text{LO}}+\ii \mathcal{A}_{\chi\chi
      h_i}^{\text{VC}}+\ii \mathcal{A}_{\chi\chi h_i}^{\text{CT}}\,,
\end{equation}
where $\ii \mathcal{A}_{\chi\chi
  h_i}^{\text{LO}}$ is the LO contribution, $\ii
\mathcal{A}_{\chi\chi h_i}^{\text{VC}}$ are the virtual vertex corrections and
$\ii \mathcal{A}_{\chi\chi h_i}^{\text{CT}}$ are the counterterms contributions. 
The LO amplitude is
\begin{equation}
    \ii \mathcal{A}_{\chi\chi h_i}^{\text{LO}} = g_{\chi\chi h_i}
    \varepsilon(p)\cdot\varepsilon^*(p) =  2 \gX \mX  
    \varepsilon(p)\cdot\varepsilon^*(p)\begin{cases}
        \sin\alpha\,, \quad i=1\\
        \cos\alpha\,, \quad i=2
    \end{cases} \,,
\end{equation}
where $p$ denotes the four-momentum of the incoming VDM particle
and $\epsilon$ its polarization vector.\\
The vertex counterterm amplitudes for $i=1,2$ are
\begin{subequations}    
    \begin{align}
        &\ii \mathcal{A}^{\text{CT}}_{\chi\rightarrow\chi h_1} =
          \left[\frac{1}{2}\cbrak{g_{\chi\chi h_2}\delta
          Z_{h_2h_1}+g_{\chi\chi h_1}\delta Z_{h_1h_1}} + g_{\chi\chi
          h_1} \delta Z_{\chi\chi}+\delta g_{\chi\chi
          h_1}\right]\varepsilon(p)\cdot\varepsilon^*(p)\\ 
        &\ii \mathcal{A}^{\text{CT}}_{\chi\rightarrow\chi h_2} =
          \left[\frac{1}{2}\cbrak{g_{\chi\chi h_1}\delta
          Z_{h_1h_2}+g_{\chi\chi h_2}\delta Z_{h_2h_2}} + g_{\chi\chi
          h_2} \delta Z_{\chi\chi}+\delta g_{\chi\chi
          h_2}\right]\varepsilon(p)\cdot\varepsilon^*(p)\,, 
    \end{align}
    \label{XXH::CT}
\end{subequations}
with the counterterms $\delta g_{\chi\chi h_i}$ ($i=1,2$) for the respective tree-level couplings
\beq
g_{\chi \chi h_1} &=& 2 \gX \mX \sin \alpha \\
g_{\chi \chi h_2} &=& 2 \gX \mX \cos \alpha 
\eeq
derived from 
\begin{equation}
    \delta g_{\chi\chi h_i}  = \sum_{p} \frac{\partial g_{\chi\chi
        h_i}}{\partial p} \delta p\,,\quad p\in \{\mX^2, \gX,\alpha\}\,.
\end{equation}
At NLO two additional tensor structures arise 
\begin{equation}
    \ii \mathcal A^{\text{NLO}} =
    \cbrak{\dots}\underbrace{\varepsilon(p_{\text{in}})\cdot\varepsilon^*(p_{\text{out}})}_{\sim
      \text{LO}} + \cbrak{\dots}\underbrace{ \cbrak{p_{\text{in}}\cdot
        \varepsilon^*(p_{\text{out}})}\cbrak{p_{\text{out}}\cdot
        \varepsilon(p_{\text{in}})}}_{\sim \text{NLO}}\,,
\end{equation}
where $p_{\text{in}}$ ($p_{\text{out}}$) is the incoming (outgoing) momentum of the DM
vector gauge boson. The additional new tensor structure (denoted by $\sim$ NLO) 
vanishes by assuming $p_{\text{in}} = p_{\text{out}}$ implying $\varepsilon(p)\cdot p =0$. 
As for the amplitude that corrects the LO contribution we have checked that it is
UV finite.
 The amplitude is then projected onto the corresponding tensor structure, the vertex corrections
are plugged in the generic diagram in \cref{VERTEXCORRECTIONS} which
contributes to the operator $\chi_{\mu}\chi^{\mu}m_q \bar q q$. This contribution
is referred to as $f_q^{\text{vertex}}$.

\subsection{Mediator Corrections}
For the mediator correction one takes the self-energy corrections to the two-point functions with all
 external Higgs fields and inserts them in the one-loop propagator in the generic amplitude in
\cref{MEDIATORCORRECTIONS}. The self-energy contribution to the $h_i h_j$
propagator ($i,j=1,2$) reads  
\begin{equation}
    \Delta_{h_ih_j} =
    -\frac{\hat{\Sigma}_{h_ih_j}(p^2=0)}{m_{h_i}^2m_{h_j}^2} \;,
\end{equation}
with the renormalised self-energy matrix
\begin{equation}
  \begin{pmatrix}
      \hat{\Sigma}_{h_1h_1}&\hat{\Sigma}_{h_1h_2}\\
      \hat{\Sigma}_{h_2h_1}&\hat{\Sigma}_{h_2h_2}
  \end{pmatrix} 
  \equiv
   \hat\Sigma(p^2) = \Sigma(p^2) - \delta m^2 -\delta T +\frac{\delta Z}{2} \cbrak{p^2-\mathcal{M}^2} + \cbrak{p^2-\mathcal{M}^2}\frac{\delta Z}{2}\,,
\end{equation}
where the mass matrix $\mathcal{M}$ and the tadpole counterterm matrix
$\delta T$ are defined in \cref{RENORM::MASSMATRIX}. The $Z$-factor
matrix $\delta Z$ corresponds to the matrix with the components
$\delta Z_{h_ih_j}$ defined in \cref{RENORM::ZFACTOR}.  
Projecting the resulting one-loop correction on the corresponding
tensor structure, we get the one-loop correction to the
Wilson coefficient of the operator $\chi_{\mu}\chi^{\mu}m_q \bar q q$ 
\begin{equation}
    f_q^{\text{med}} = \frac{g \gX \mX}{2m_W} \sum_{i,j} R_{\alpha,i2}R_{\alpha,j1}\Delta_{h_ih_j}\,,
\end{equation} 
with the rotation matrix $R_{\alpha}$ defined in \cref{VDM::massbasis}.

\subsection{Box Corrections \label{sec:boxcorrections}}
In the following we want to present the treatment of the box diagrams
contributing to the SI cross section. 
The relevant terms of the box diagram tensor structures in the  
spin-independent cross section are extracted using an expansion in the loop diagrams.
This expansion is performed in terms of the non-relativistic momentum  $p_q$  of the external quark~\cite{Abe:2018emu}.
The box diagrams contribute to 
$ X_{\mu} X^{\mu} m_q \bar qq $ and the twist-2 operators which becomes clear if we write~\cite{Ertas:2019dew,Hisano:2010ct,Hisano:2015bma}
\begin{equation}
    \bar q\ii \partial_{\mu}\gamma_{\nu}q = \mathcal{O}_{\mu\nu}^q + \bar q\frac{\ii \partial_{\mu}\gamma_{\nu}-\ii \partial_{\nu}\gamma_{\mu}}{2} q + \frac{1}{4} g_{\mu\nu} m_q \bar qq\,,
    \label{BOX::rewrite}
\end{equation}
where the asymmetric part in 
\cref{BOX::rewrite} does not contribute to the SI cross section. We will refer to these one-loop contributions to
the corresponding tree-level Wilson coefficients as $f_q^{\text{box}}$
and $g_q^{\text{box}}$. 

There are still contributions from the effective gluon interaction with
the DM particles that will contribute to the Wilson coefficient $f_G$ in Eq.~(\ref{tree::effoperators}b). \color{black}  
As shown in  \cite{Abe:2018emu} \color{black} the use of \cref{tree::gluonmap}
to obtain the gluon interaction induces large errors. An ansatz
was proposed in Ref.~\cite{Ertas:2019dew}   for 
heavy quarks compared to the mediator mass, by deriving an effective coupling between
two Higgs bosons and the gluon fields. Integrating out the
top-quark leads to the effective two-Higgs-two-gluon
coupling\cite{Ertas:2019dew}
\begin{equation}
    \mathcal L^{hhGG} = \frac{1}{2} d_G^{\text{eff}} h_i h_j
    \frac{\alpha_S}{12\pi} G^a_{\mu\nu}G^{a \,\mu\nu}\,, 
    \label{box::dgeff}
\end{equation}
with
\begin{equation}
    d_G^{\text{eff}} \to \cbrak{d_G^{\text{eff}}}_{ij} =
    (R_\alpha)_{i1} (R_\alpha)_{j1} \frac{1}{v^2}\, .
\end{equation}
We note that In Ref.~\cite{Ertas:2019dew}, the full two-loop calculation was
performed showing very good agreement with the approximate result
for mediator masses below $m_t$.
Moreover, the box contribution to the NLO SI direct detection
cross section is several orders of magnitude below the LO contribution as we will show later.
\begin{figure}
    \centering
\hspace*{-4.5cm}
    \subfigure{\includegraphics[width = 0.20\textwidth]{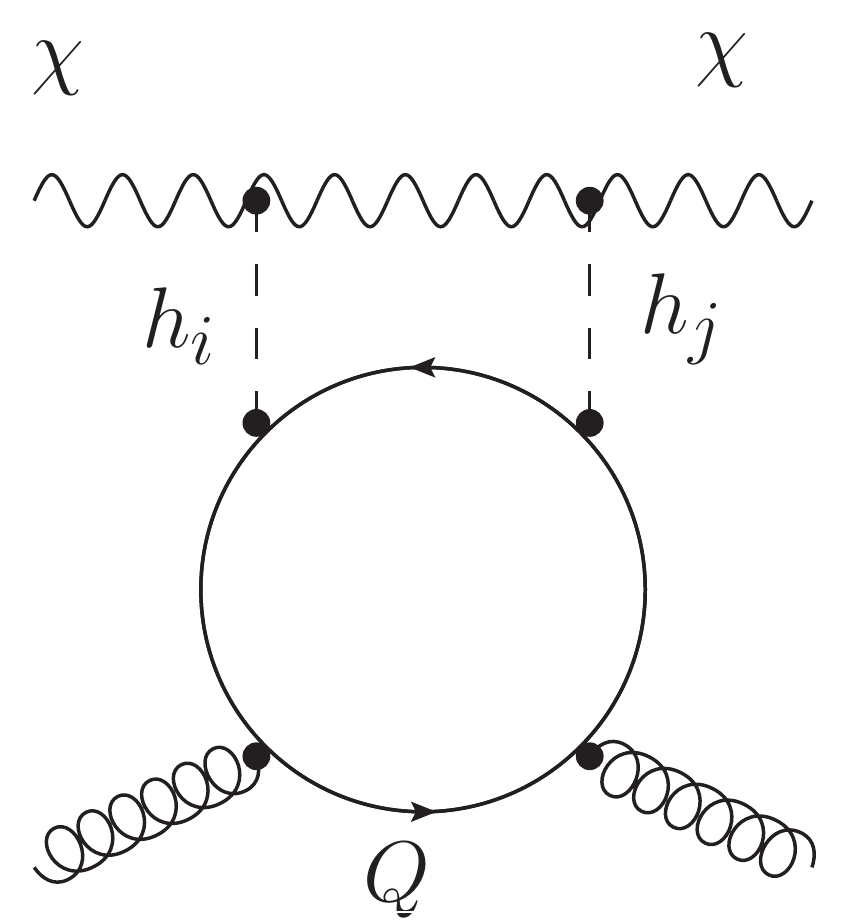}}
\\[-3cm]
\hspace*{6cm}
    \subfigure{\includegraphics[width =
      0.25\textwidth]{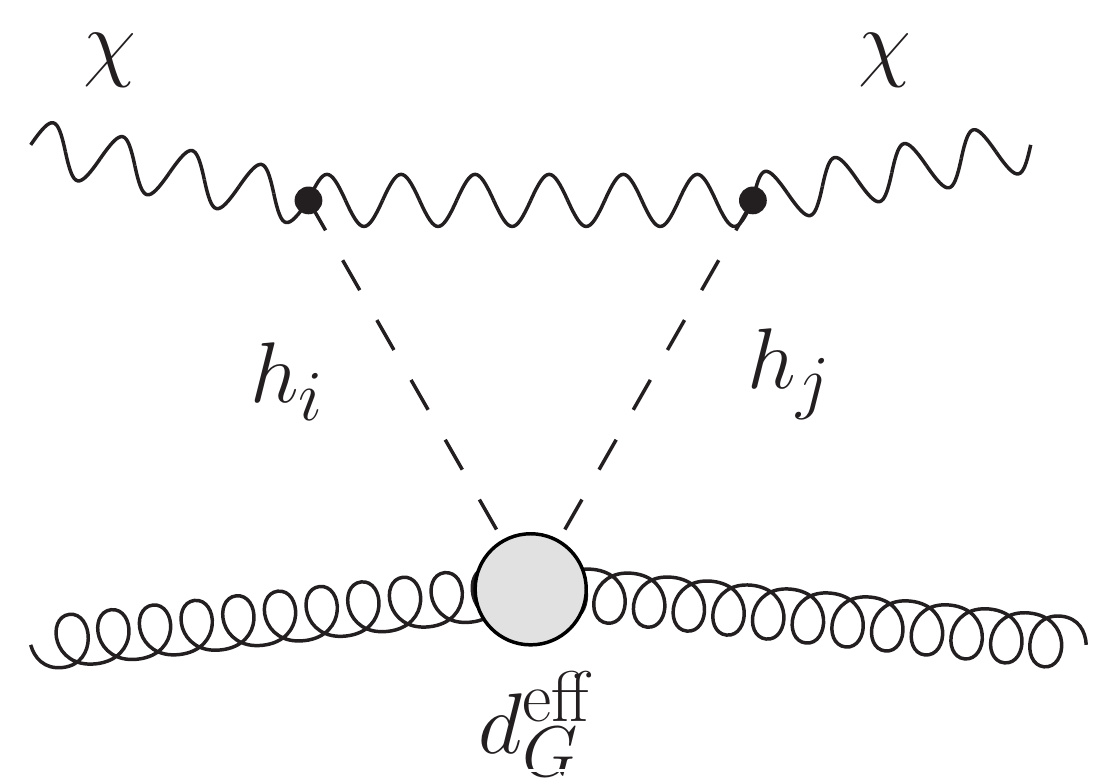}}
\vspace*{0.3cm}
    \caption{The full two-loop gluon interaction with the DM candidate
      (left) and the effective two-loop interaction after integration
      out the heavy quarks (right). \label{BOX::GLU}} 
\end{figure}

The diagram in \cref{BOX::GLU} (right) yields the following contribution to the
Lagrangian 
\begin{equation}
  \mathcal{L}_{\text{eff}}\supset  
\left( d_G^{\text{eff}}\right)_{ij} C^{ij}_\triangle \chi_{\mu}\chi^{\mu} \frac{-\alpha_S}{12\pi}
  G^a_{\mu\nu}G^{a \, \mu\nu}\,, 
\end{equation} 
where $C^{ij}_\triangle$ denotes the contribution from the triangle
loop built up by $h_i$, $h_j$ and the VDM particle. It has to be
extracted from the calculated amplitude of 
\cref{BOX::GLU} (right). Using \cref{eq:ggxx}  the contributions by
the box topology to the gluon-DM interaction is given by
\begin{equation}
    f_G^{\text{top}}= \left( d_G^{\text{eff}} \right)_{ij} C^{ij}_\triangle \frac{-\alpha_S}{12 \pi} \,.
\end{equation}

\subsection{The SI One-Loop Cross Section}
We can now write the NLO EW SI cross section using the effective one-loop form factor 
\begin{equation}
    \frac{f_N^{\text{NLO}}}{m_N} = \sum_{q=u,d,s} f_q^{\text{NLO}}
    f^N_{T_q} + \sum_{q=u,d,s,c,b}\frac{3}{4} \cbrak{q(2)+\bar q(2)}
    g_q^{\text{NLO}} - \frac{8\pi}{9\alpha_S} f^N_{T_G}
    f_G^{\text{NLO}}\,, 
\label{eq:sigmanlo}
\end{equation}
with the Wilson coefficients at one-loop level given by
\begin{subequations}
    \begin{align}
    &f_q^{\text{NLO}} =
      f_q^{\text{vertex}}+f_q^{\text{med}}+f_q^{\text{box}}
      \\
    &g_q^{\text{NLO}} = g_q^{box} \\
    &f_G^{\text{NLO}} = -\frac{\alpha_S}{12\pi}
      \sum_{q=c,b,t}\cbrak{f_q^{\text{vertex}}+f_q^{\text{med}}} +
      f_G^{\text{top}}\,. 
\end{align}
\end{subequations}
Like at LO, the heavy quark contributions of
  $f_q^{\text{vertex}}$ and $f_q^{\text{med}}$ have to be attributed
  to the effective gluon interaction. With the LO form factor given by
\beq
\frac{f_N^{\text{LO}}}{m_N} = f_q^{\text{LO}} \left[\sum_{q=u,d,s} f^N_{T_q}
    + \sum_{q=c,b,t} \frac{2}{27} f_{T_G}^N \right] \,,
\eeq
where $f_q^{\text{LO}}$ has been given in
Eq.~(\ref{eq:fqlo}) \color{black}, we have for the NLO EW SI cross section at leading
order in $\alpha_S$,
\beq
\sigma_N = \frac{1}{\pi} \left( \frac{m_N}{\mX + m_N}\right)^2
\left[ |f_N^{\text{LO}}|^2 + 2 \mbox{Re} \left( f_N^{\text{LO}}
  f_N^{\text{NLO}*} \right) \right]\;.
\eeq

\subsection{The inclusion of the lower vertex corrections \label{sec:lvcorrections}}
In our approach in~\cite{Glaus:2019itb} we have neglected the EW
corrections of the lower vertex $q\overline{q}  h_i$ due to the
missing cancellation of IR divergencies. This naive approach gives rise to several subtle problems
to be discussed in the following. The field strength renormalisation constants $\delta Z_{h_ih_j}$ in ~\cref{RENORM::ZFACTOR} for the Higgs boson mediator are introduced artificially. Considering the full process these \textit{internal} field strength renormalisation constants (referred to as $\mathcal{A}^{\delta Z}_i$) would cancel exactly
\begin{equation}
    \left(\mathcal{A}_{u p V}^{\mathrm{VC}}+\mathcal{A}_{u p V}^{\mathrm{gCT}}\right)+\left(\mathcal{A}_{m e d}^{\mathrm{VC}}+\mathcal{A}_{m e d}^{\mathrm{gCT}}\right)+\left(\mathcal{A}_{L V}^{\mathrm{VC}}+\mathcal{A}_{L V}^{\mathrm{gCT}}\right)+\underbrace{\mathcal{A}_{u p V}^{\delta Z}+\mathcal{A}_{m e d}^{\delta Z}+\mathcal{A}_{L V}^{\delta Z}}_{=0} = \mathrm{UV~finite}\,,
    \label{cancellation}
\end{equation}
where $\mathcal{A}^{\text{VC}}_i$ corresponds to the genuine one-loop diagrams and $\mathcal{A}^{\text{gCT}}_i$ to the counterterm amplitude (without the $\delta Z_{h_ih_j}$ factor). The box contributions are not relevant for the problem and therefore dropped in the following discussion. The contributions of the upper vertex $\chi\chi h_i$ are referred to as $UpV$, the lower vertex $q\bar q h_i$ as $LoV$  and mediator corrections as $med$, respectively. The artificial introduction of the $\delta Z$ part allows to cancel the UV-poles topology-wise
\begin{equation}
\left(\mathcal{A}_{i}^{\mathrm{VC}}+\mathcal{A}_{i}^{\mathrm{gCT}}\right)_{\text{fin.}}+\underbrace{\left(\mathcal{A}_{i}^{\mathrm{VC}} +\mathcal{A}_{i}^{\mathrm{gCT}}\right)_{\Delta}+\left(\mathcal{A}_{i}^{\delta Z}\right)_{\Delta}}_{=0}+\left(\mathcal{A}_{i}^{\delta Z}\right)_{\text{fin.}}\,,\quad (i=UpV, LoV, med)\,,
\end{equation}
where $\cbrak{\dots}_{\Delta}$ indicates the explicit UV pole of the amplitude and $\cbrak{\dots}_{\text{fin.}}$ the finite part, respectively. By dropping the full lower vertex in the matching of the Wilson coefficients the finite piece $-\mathcal{A}_{LV}^{\delta Z}$ would remain in the amplitude due to the missing cancellation indicated in \cref{cancellation}. \\
As an additional issue, the chosen renormalisation scheme for the
mixing angle $\delta \alpha$ was shown~\cite{Denner:2018opp,
  Denner:2019xti} to be numerically stable only if either $\delta
\alpha$ and $\delta Z_{h_ih_j}$ occur in a specific combination, or by
including \textit{all diagrams} of the process yielding a numerically
stable combination of $\delta \alpha$. The former case is present in
on-shell decays and the latter is our present approach. This numerical
instability is related to the $\frac{1}{m_{h_1}^2-m_{h_2}^2}$ mass
pole used in the definition of the mixing angle. Note that this pole
also occurs explicitly in the off-diagonal $\delta Z_{h_ih_j}$ field
strength renormalisation constants of the Higgs bosons. By dropping
the lower vertex the cancellation of the mass pole due to the mixing
angle counterterm combination would also not be present anymore and
the result would be numerically unstable. Note that by stability we mean 
that the numbers are not unnaturally large.

\begin{figure}
    \centering
\hspace*{-5cm}
    \subfigure{\includegraphics[width = 0.25\textwidth]{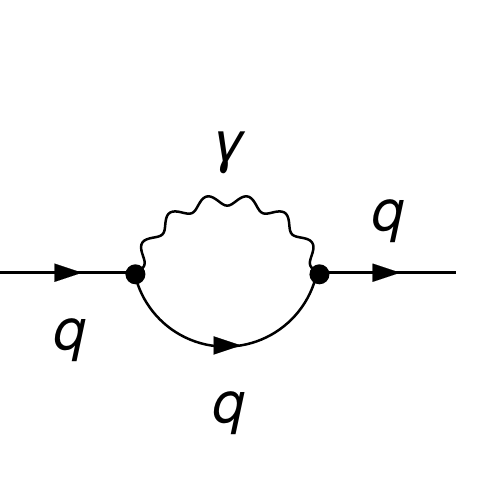}}
\\[-4cm]
\hspace*{6cm}
    \subfigure{\includegraphics[width =
      0.25\textwidth]{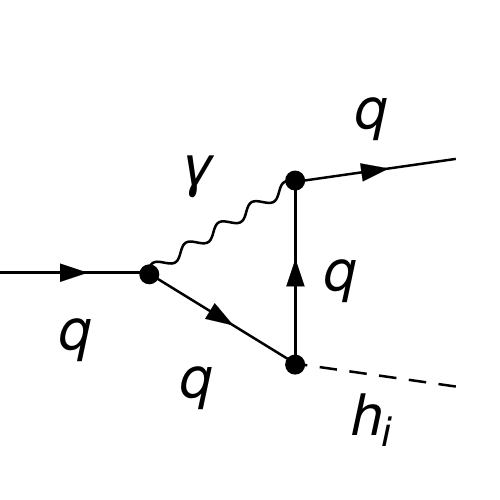}}
\vspace*{0.3cm}
    \caption{QED subset contribution to the lower vertex
      coefficient. The QED subset is UV finite and contains all IR
      divergences. \label{QEDsubset} } 
\end{figure}

By artificially including the $\delta Z_{h_ih_j}$ in the
mediator counterterm and dropping the lower vertex as proposed in \cite{Glaus:2019itb}, the Higgs mediator
is treated as \textit{on shell} so that the mass pole $1/m_h^2-m_{\phi}^2$ of the $\delta Z$ remainder ($-\mathcal{A}_{LV}^{\delta Z}$ in \cref{cancellation}) and the $\delta \alpha$ mass pole remainder of the upper vertex cancel each other exactly. In this way numerically stable EW NLO corrections could be obtained. 
However, this cancellation is unphysical and should be
avoided, indicating that the dropping of the lower vertex is not the
optimal solution.

The IR divergent diagrams of the lower vertex, shown in
\cref{QEDsubset}, form a UV-finite subset, referred to as QED subset
in the following. This QED subset includes all lower vertex
corrections and quark self-energies containing a photon
line. Expanding this QED subset for strictly vanishing external quark
momentum and neglecting all terms of the order $\mathcal{O}(p_Q^2)$
which is also compatible with the expansion used in the box calculation in~\cite{Glaus:2019itb}, allows to regulate all IR divergencies. In this way the lower vertex, which is a sum of all contributions without a photon and the QED subset, is explicitly UV finite and all IR divergencies are regulated by the strict vanishing quark momentum expansion. 
Using the expansion allows to include the lower vertex in the matching of the Wilson coefficients without including any IR divergences and thereby the cancellation of \cref{cancellation} is present. The unphysical treatment of the internal Higgs mediator is avoided and a numerically stable result is obtained. 

The inclusion of the EW corrections to
the lower vertex, however, invalidate the replacement rule of
\cref{tree::gluonmap} for the lower vertex.
 So far the QCD trace anomaly is calculated at one-loop (QCD) level to find the relation between the heavy quark operator $m_Q \overline{Q}Q\, (Q=c,b,t)$ and gluon field-strength operator $G_a^{\mu\nu}G^{a,\mu\nu}$. Including EW corrections spoil this replacement rule and the proper matching is beyond the scope of this analysis. Therefore, the EW corrections of the lower vertex with a heavy quark cannot be considered in the calculation of the gluon contribution to the spin-independent cross section, since otherwise the cancellation of \cref{cancellation} would fail again. 
Hence, the DM-gluon interaction is determined only considering EW LO
diagrams and using the replacement rule \cref{tree::gluonmap}. In this way the problem discussed above with the $\delta Z_{h_ih_j}$ is avoided.

\section{Numerical Analysis \label{sec:result}}

The VDM model was implemented in the \texttt{ScannerS}~\cite{Coimbra:2013qq, Costa:2015llh} code 
which automatises the parameter scan. The points generated are constrained by
\begin{itemize}

\item 
The SM-like Higgs boson has a mass of $m_h=125.09$~GeV~\cite{Aad:2015zhl}. 

\item
The potential is in a global minimum and all points satisfy the
theoretical constraints of boundedness from below and perturbative
unitarity. We furthermore impose the perturbativity constraint $g_\chi^2 < 4 \pi$. 

\item
The mixing angle $\alpha$ is constrained by the combined values for the
signal strengths~\cite{Aad:2015zhl}. An interface with
\texttt{HiggsBounds}~\cite{Bechtle:2008jh,Bechtle:2011sb,Bechtle:2013wla}
allows to check for collider bounds from LEP, Tevatron and
the LHC. We require agreement with the exclusion limits derived for
the non-SM-like Higgs boson at 95\% confidence level. 
The most stringent bound arises from searches for heavy $ZZ$ resonances~\cite{Aaboud:2017rel}.  

\item
Calculations of the Higgs decay widths and branching ratios are performed with {\tt
  sHDECAY}~\cite{Costa:2015llh}\footnote{The program {\tt sHDECAY} can
  be downloaded from the url:
  \url{http://www.itp.kit.edu/~maggie/sHDECAY}.}, which includes the
state-of-the-art higher-order QCD corrections. The code {\tt sHDECAY}
is based on the implementation of the models in {\tt
  HDECAY}~\cite{Djouadi:1997yw,Djouadi:2018xqq}. 

\item
Information on the DM particle is taken into account from LHC searches through the invisible width of the SM Higgs 
boson~\cite{Bechtle:2008jh,Bechtle:2011sb,Bechtle:2013wla}.

\item
The DM relic abundance was calculated with
\texttt{MicrOMEGAs}~\cite{Belanger:2006is,Belanger:2007zz,Belanger:2010pz,Belanger:2013oya},
and compared with the current experimental result from the Planck
Collaboration~\cite{Ade:2015xua}, 
\beq
({\Omega}h^2)^{\rm obs}_{\rm DM} = 0.1186 \pm 0.002 \;. 
\eeq
We require the calculated abundance to be equal
to or smaller than the observed one, that is, we allow the DM  not to
saturate the relic density and therefore define a DM fraction 
\begin{eqnarray}
f_{\chi\chi} = \frac{({\Omega} h^2)_{\chi}}{(\Omega h^2)^{\rm
  obs}_{\text{DM}}}\,, 
\label{eq:dmfraction}
\end{eqnarray} 
where $(\Omega h^2)_{\chi}$ stands for the calculated DM relic
abundance of the VDM model. 

\item
DM indirect detection does not play a relevant role here.
See~\cite{Glaus:2019itb} for details. 

\item
The sample was generated taking into account the direct detection
bound on the DM nucleon SI cross section at LO. The most stringent experimental bound is
the one from the XENON1T~\cite{Aprile:2017iyp,Aprile:2018dbl} experiment. We apply the
latest XENON1T upper bounds~\cite{Aprile:2018dbl} for a DM mass above
6~GeV and the combined limits from CRESST-II~\cite{Angloher:2015ewa} and 
 CDMSlite~\cite{Agnese:2015nto} are used for lighter DM particles. 
Because the experimental limits on the DM-nucleon scattering assume 
the DM candidate to make up for all of the DM
  abundance, the correct quantity to be compared
  with the experimental limits is the effective DM-nucleon
  cross-section defined by
\beq 
\sigma^{\rm eff}_{\chi N} \equiv f_{\chi\chi} \sigma_{\chi N} \;,
\label{cor_factor}
\eeq
where $\chi N$ stands for the scattering of the VDM particle $\chi$ with the
nucleon $N$, and $f_{\chi\chi}$ denotes the respective DM fraction,
defined in Eq.~(\ref{eq:dmfraction}) \color{black}. In our numerical analysis, we
use the LO and NLO results for $N=p$. 
\end{itemize}

The ranges of the input parameters for the scan are shown in Table~\ref{tab:vdmscan}\color{black}. From
now on we denote the non-SM like Higgs boson mass as $m_\phi$ and the SM-like Higgs boson mass as $m_h$. 
\begin{table}
\begin{center}
\begin{tabular}{l|cccc} \toprule
& $m_{\phi}$ [GeV] & $\mX$ [GeV] & $v_S$ [GeV] & $\alpha$ \\ \hline
min & 1 & 1 & 1 & $-\frac{\pi}{4}$ \\
max & 1000 & 1000 & $10^7$ & $\frac{\pi}{4}$ \\ \bottomrule
\end{tabular}
\caption{Input parameters for the VDM model scan, all parameters
  varied independently between the given minimum and maximum
  values. The SM-like Higgs boson mass is set
  $m_h=125.09$~GeV and the SM VEV
  $v=246.22$~GeV. 
  \label{tab:vdmscan}}
\end{center}
\end{table}
The remaining input parameters, gauge, lepton and quark masses,
electric coupling, Weinberg angle and weak $SU(2)$ coupling, are set to
\beq
\begin{array}{lllllllll}
m_W &=& 80.398 \mbox{ GeV } \;, \quad & m_Z &=& 91.1876 \mbox{ GeV }
\;, \quad & \sin\theta_W &=& 0.4719 \;,\\
m_e &=& 0.511\cdot 10^{-3} \mbox{ GeV } \;, \quad & m_\mu &=& 0.1057 \mbox{
GeV } \;, \quad & m_\tau &=& 1.777 \mbox{ GeV } \;, \\
m_u &=& 0.19 \mbox{ MeV } \;, \quad & m_d &=& 0.19 \mbox{ MeV } \;, \quad 
& m_s &=& 0.19 \mbox{ MeV } \;, \\
m_c &=& 1.4 \mbox{ GeV } \;, \quad & m_b &=& 4.75 \mbox{ GeV } \;,
\quad & m_t &=& 172.5 \mbox{ GeV } \;. 
\end{array}
\eeq
For the proton mass we take \beq
m_p = 0.93827 \mbox{ GeV}\,.
\eeq

\section{Results}

We will now present the results with the NLO corrections, focusing on the 
main changes relative to our previous work~\cite{Glaus:2019itb}. The sample
used complies with all theoretical and experimental bounds described in the 
previous section. We note that the bound for direct detection at LO is already
imposed and that, in order to be able to  compare with the Xenon limit, we applied the correction factor
  $f_{\chi\chi}$ to the LO and NLO direct detection cross section, {\it
    cf.}~Eq.~(\ref{cor_factor}\color{black}).

\subsection{Relative size of one-loop corrections}

\begin{figure}[h!]
    \centering
    \includegraphics[width=0.48\textwidth]{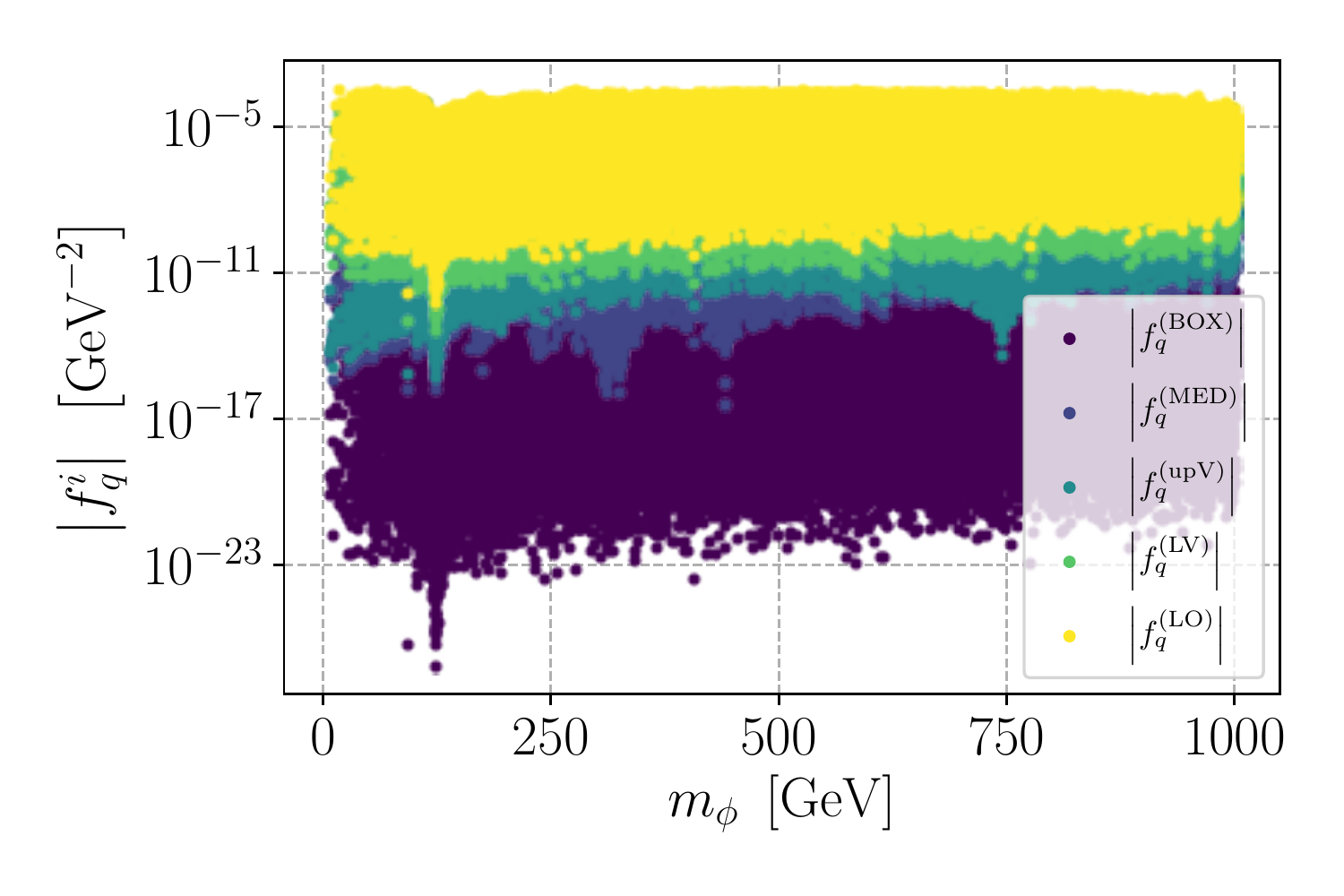}
    \includegraphics[width=0.48\textwidth]{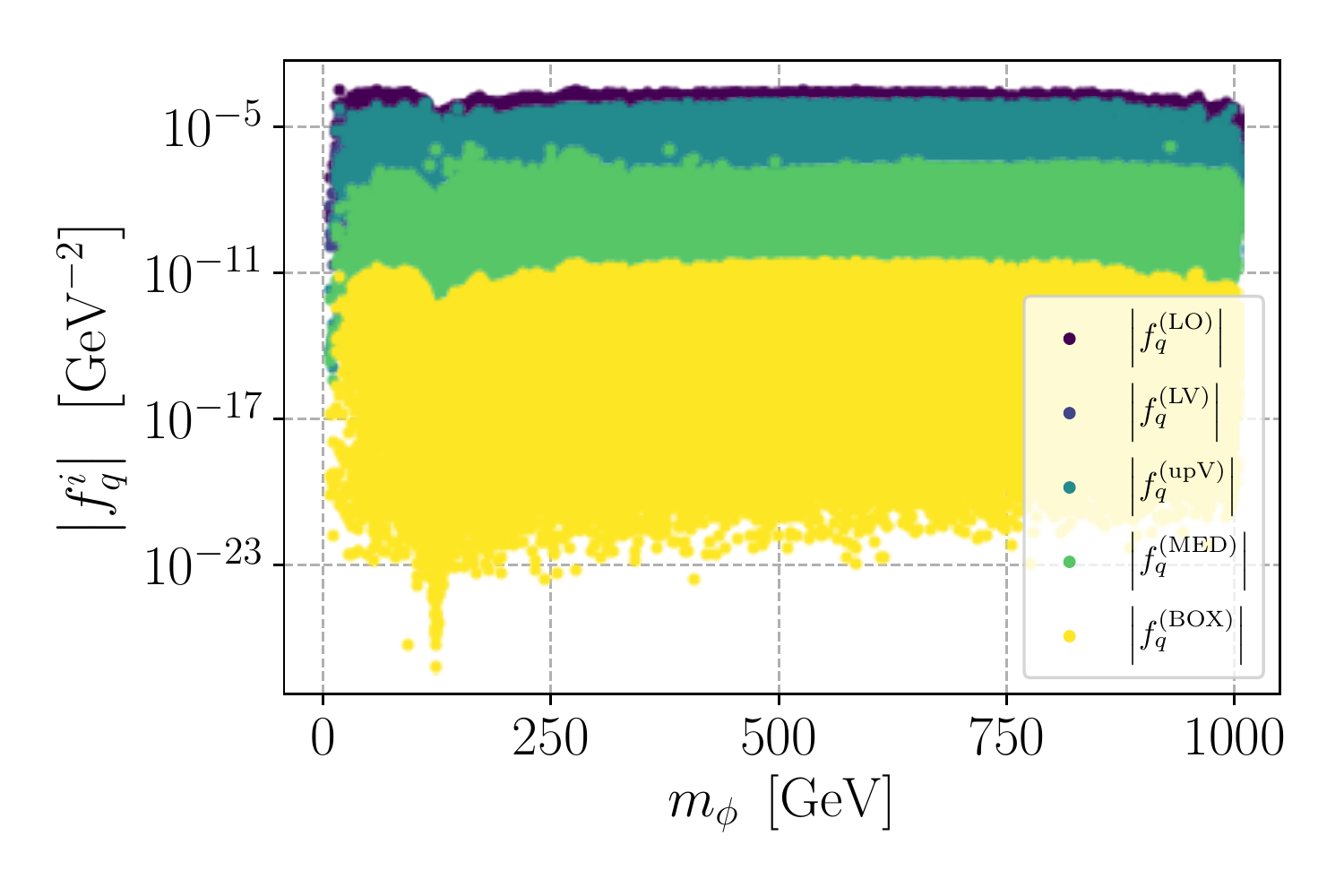}
    \caption{\textbf{we need units on the y axis which I believe are $[\sigma]/[m]$}\\Value of the Wilson coefficients of the different NLO contributions as compared to the LO,  with the LO colour on top (left)
      and reversed colour code (right).}
    \label{fig:result1}
\end{figure}
In~\cref{fig:result1} we present the values of the Wilson coefficients contributing to the LO and to the NLO cross sections with the
colour code where the largest contributions are on top (on the left)
and the inverted colour code (on the right) as a
  function of the non-SM-like Higgs boson mass $m_\phi$. The order of the relevance of the contributions
is clear from the two plots. The LO is about one order of magnitude above the most relevant one-loop corrections which are
the vertex contribution, both the lower and upper one. Another clear point revealed by the plots is that the box contributions are several orders of magnitude
below the vertex corrections and are therefore negligible.
\begin{figure}[h!]
    \centering
    \includegraphics[width=0.48\textwidth]{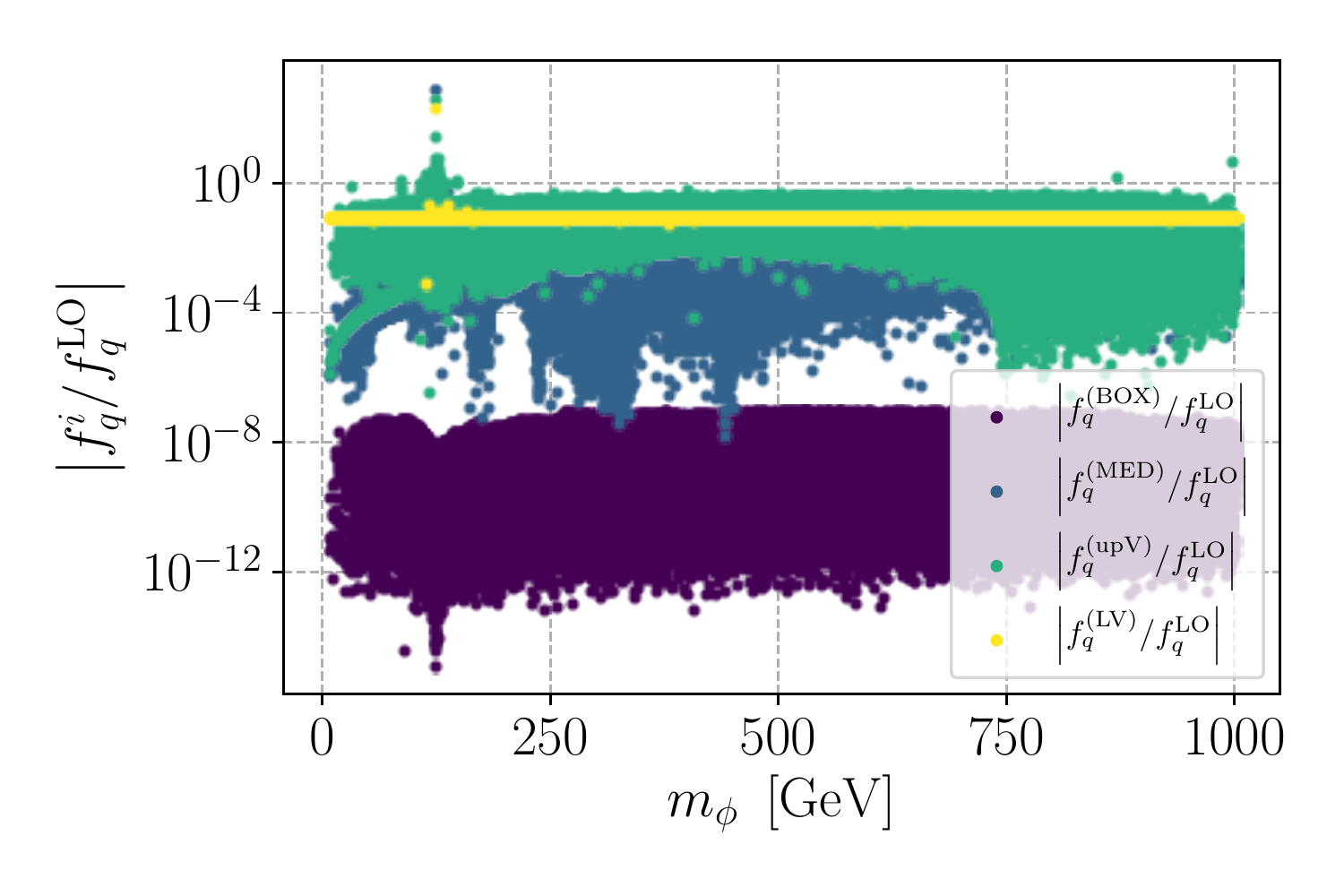}
    \includegraphics[width=0.48\textwidth]{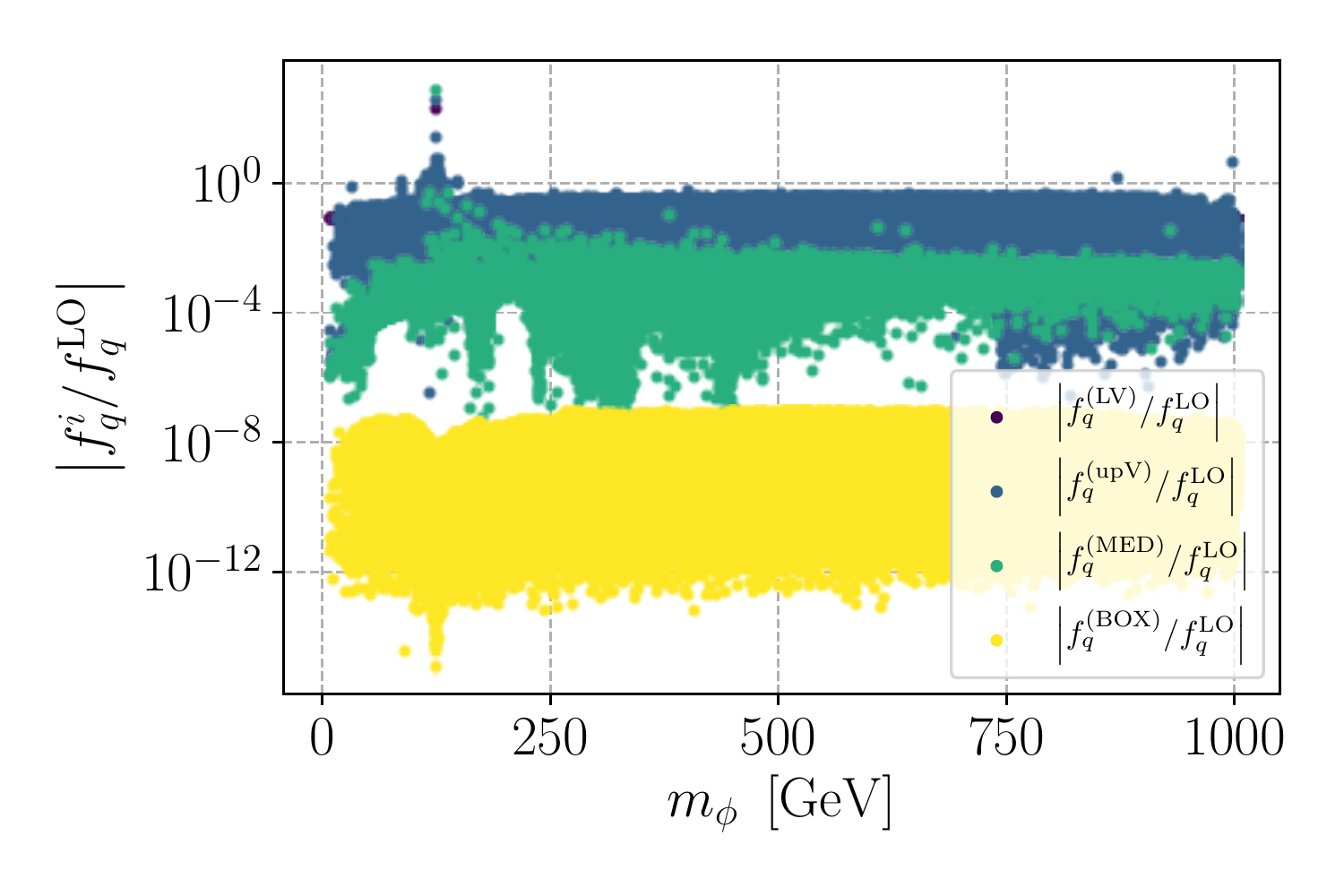}
    \caption{Value of the Wilson coefficients of the different NLO contributions, normalised to the LO coefficient,  with the LO colour on top (left)
      and reversed colour code (right).}
    \label{fig:result2}
\end{figure}
In~\cref{fig:result2} we present the same NLO Wilson coefficients but
now normalised to the LO result and the same colour code as a
  function of the non-SM-like Higgs boson mass $m_\phi$. In this plot
the relative importance of the lower and the upper vertex becomes clearer from the plot on the left. The right plot shows that the mediator
contribution also plays a role  in particular close to the
SM Higgs boson
mass. Again, box contributions are clearly
negligible. 
%

\subsection{$K$-factors and Impact of the NLO Corrections on the Xenon Limit}
\begin{figure}[h!]
    \centering
    \includegraphics[width=0.48\textwidth]{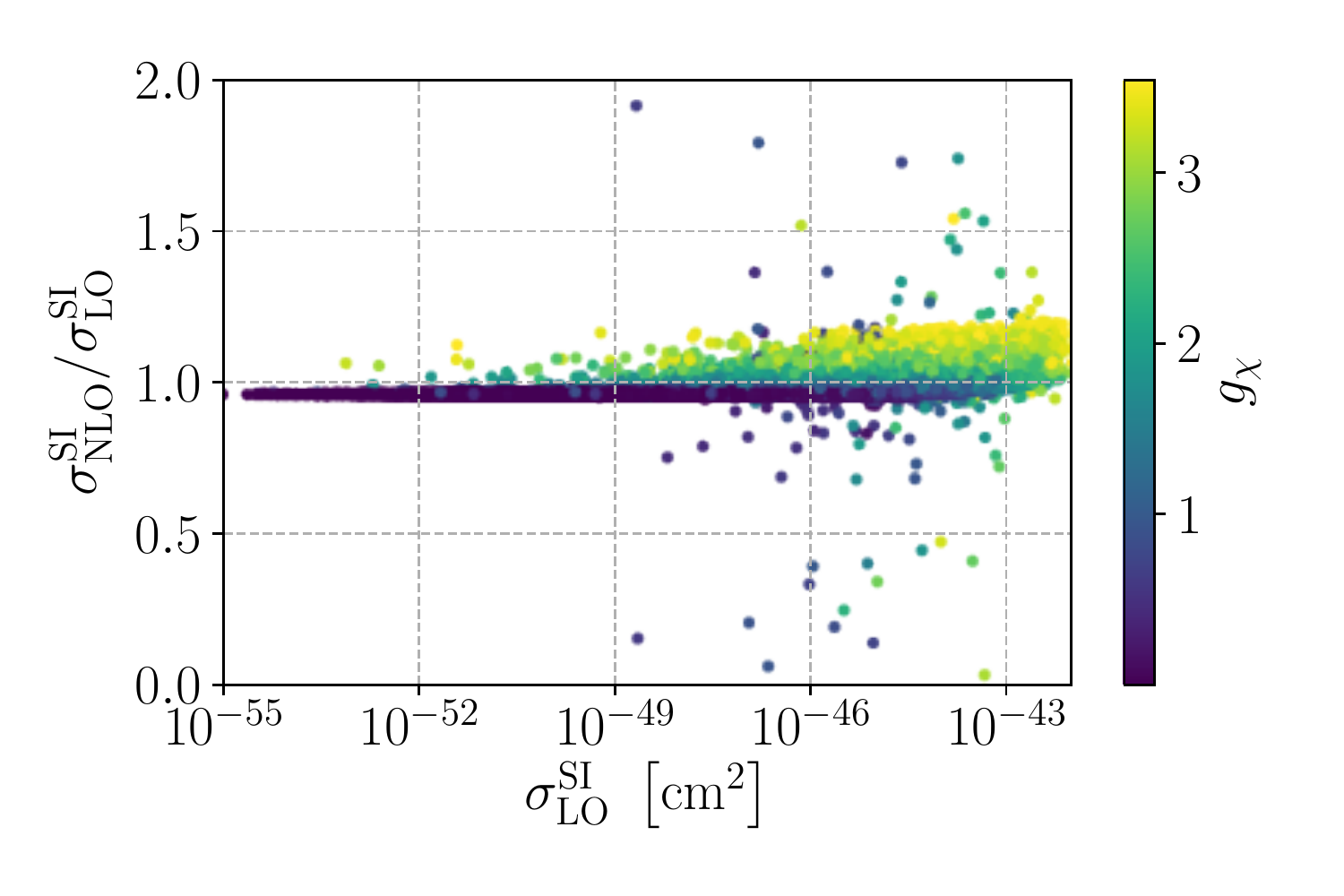}
    \includegraphics[width=0.48\textwidth]{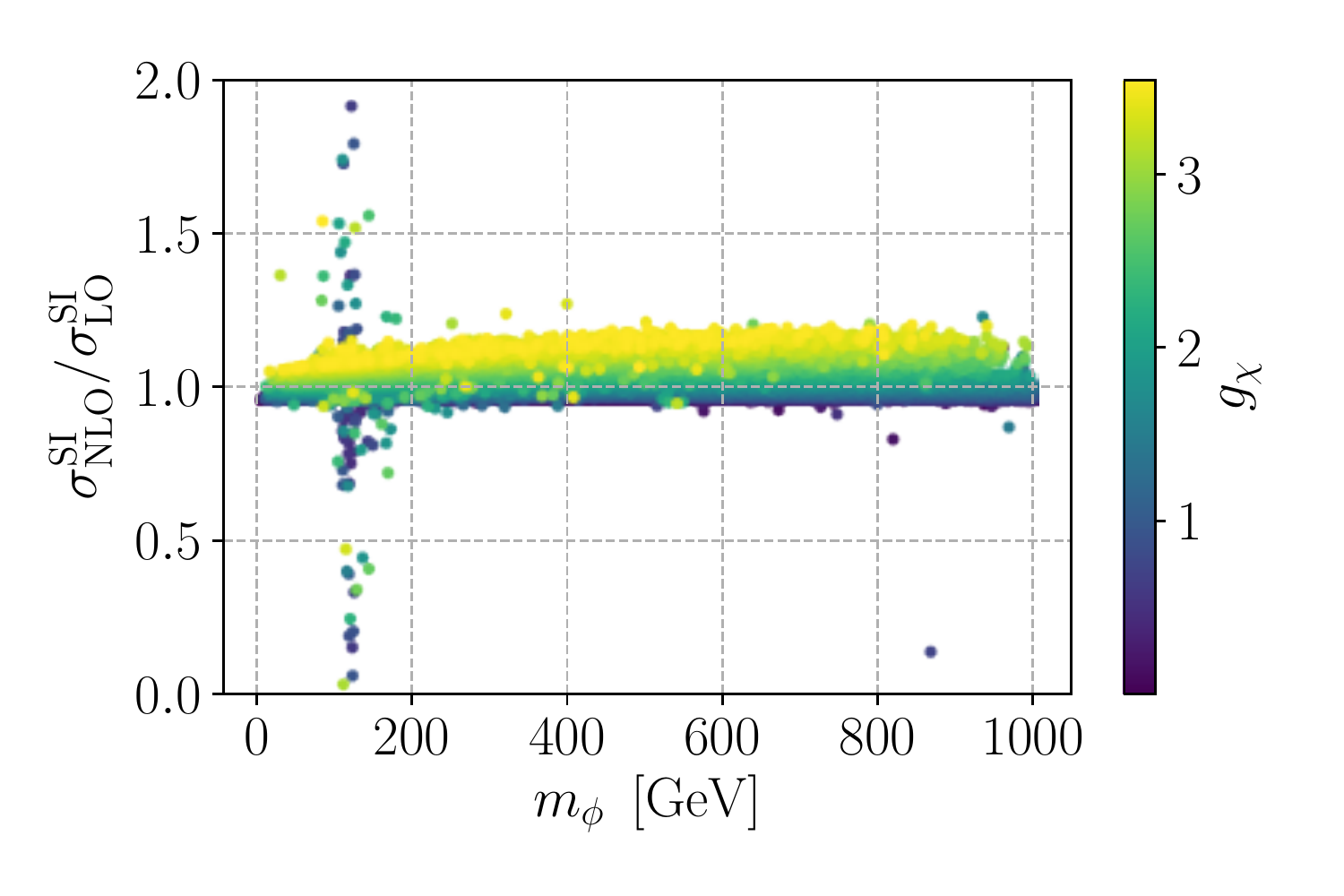}
    \caption{$K$-factor as function of the LO direct detection cross
      section (left) and as a function of the non-125 GeV Higgs mass (right). The color code 
      denotes the size of the dark gauge coupling $\gX$. }
    \label{fig:result3}
\end{figure}

We now turn to the comparison of the NLO to the LO cross section of direct detection.
In~\cref{fig:result3} we show the $K$-factor, {\it i.e.}~the ratio between NLO and LO cross section,
as a function of the LO
SI direct detection cross section (left) and as a function of the
non-SM-like Higgs boson mass $m_\phi$ (right). 
The size of $\gX$ is indicated by the color code. The main points to note are the following:
the  $K$-factor increases with $\gX$ but except for the outliers the increase is always 
below about 30\%; the outliers, clearly seen on the right plot, appear close to $m_\phi = m_h$
with $K$-factors close to 2 which are due to the resonant behaviour in the vertex contributions.  
The values for the $K$-factors are much smaller than the ones obtained in our previous study~\cite{Glaus:2019itb} 
where we could see  $K$-factors reaching 100\%. This is the main difference we found after the inclusion
of the lower vertex contribution. Also the resonant contributions are more stable with values of $K$-factors
below about 2.
The $K$-factor shows no particularly interesting dependence on the other free parameters, 
the mixing angle $\alpha$ and the vector DM mass.

Both the LO and the NLO contributions to the SI direct detection cross section are proportional to the LO amplitude and are therefore
 proportional to $\sin 2 \alpha$ and  $m_h^2 - m_\chi^2$. Hence, blind spots are the same at LO and at NLO. In our scan we did not find any other points 
 where a specific parameter combination would lead to an accidental suppression at LO that is removed at NLO. The blind spot at $\alpha = 0$ represents a 
 scenario where the SM-like Higgs boson has exactly SM-like couplings and the new scalar only couples to the Higgs and to dark matter. 
 The SM-like Higgs decouples from dark matter and we may end up with two dark matter candidates with the second scalar being metastable.

\begin{figure}[h!]
    \centering
    \includegraphics[width=0.48\textwidth]{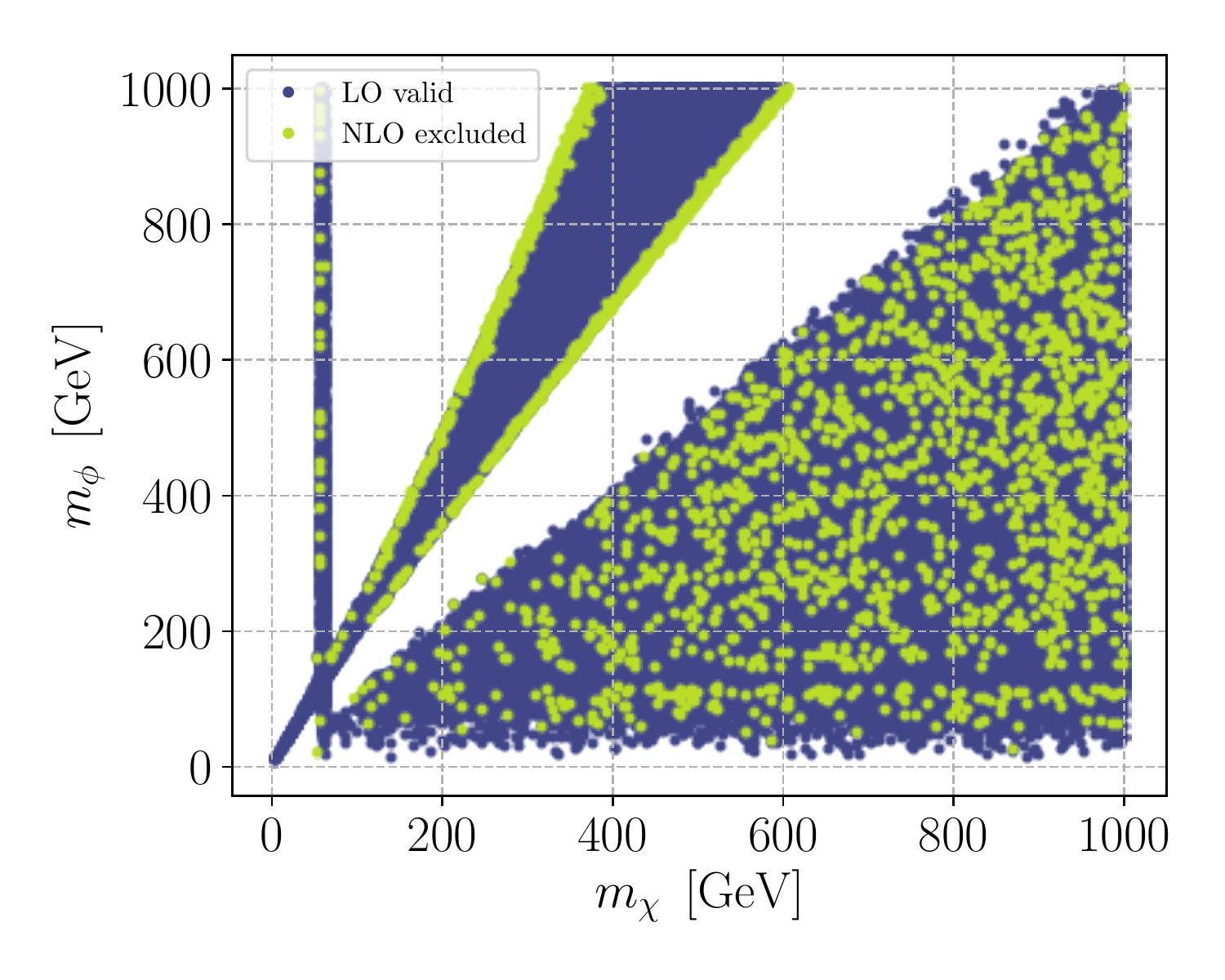}
    \includegraphics[width=0.48\textwidth]{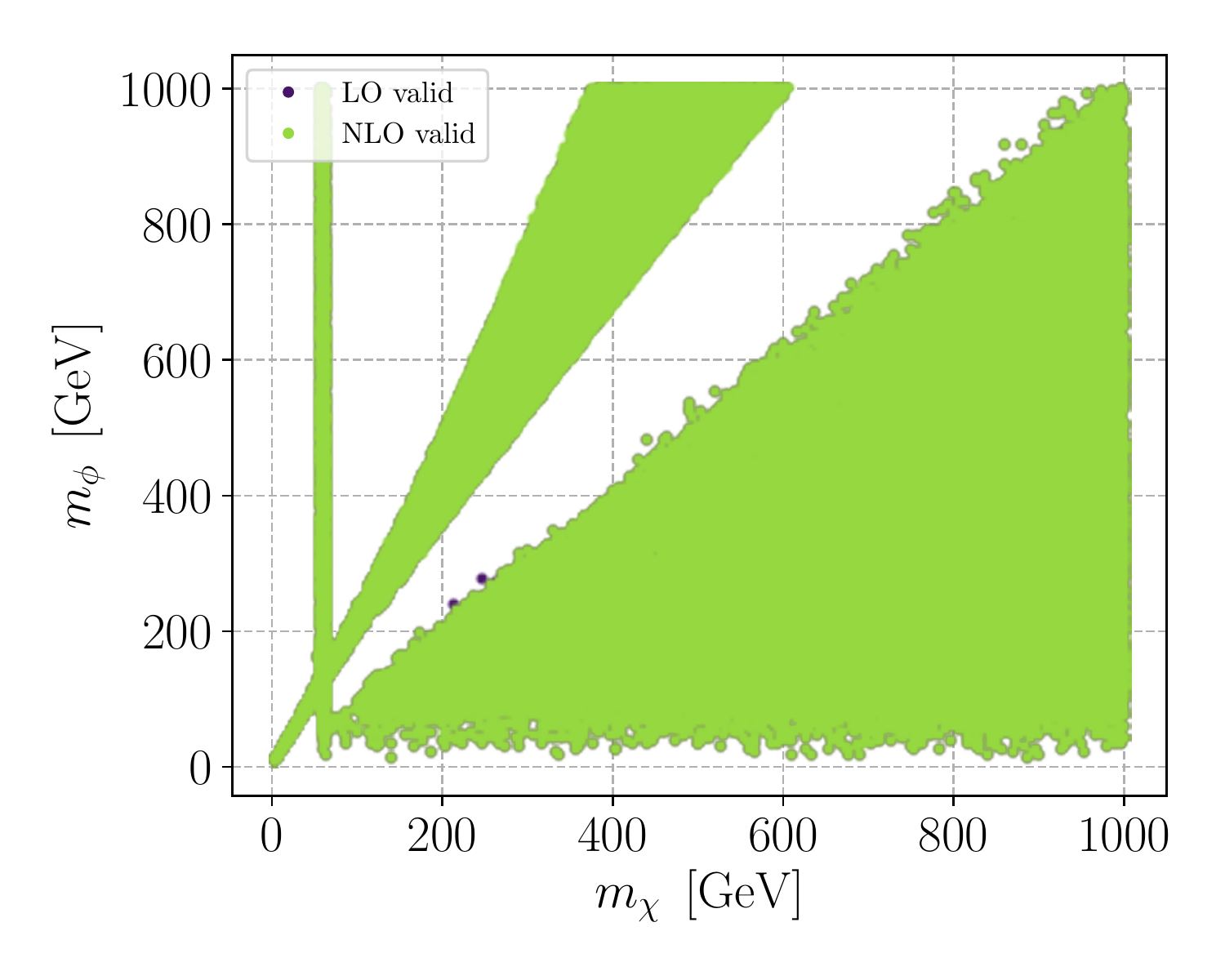}
    \caption{The SI cross section including the correction factor
      $f_{\chi\chi}$ at LO (biue) and NLO (orange) compared to the
      Xenon limit (blue-dashed) versus the DM mass $\mX$. The definition
    of the parameter sample included in the left and right plots is
    described in the text.}
    \label{fig:result4}
\end{figure}

We end this section with a discussion of the phenomenological impact of our
NLO results on the Xenon limit. In~\cref{fig:result4} we show the allowed parameter
space in the $(m_\phi, \, m_\chi)$ plane with all constraints taken into account.
The blue points are the ones valid for the LO direct detection cross section. In the left plot the green points 
are the ones excluded at NLO and in the right plot they represent the allowed points at NLO. The plots tell us that
although we see a very large number of points excluded at NLO, the difference between LO and NLO would hardly be seen in a scan.
\begin{figure}[h!]
    \centering
    \includegraphics[width=0.58\textwidth]{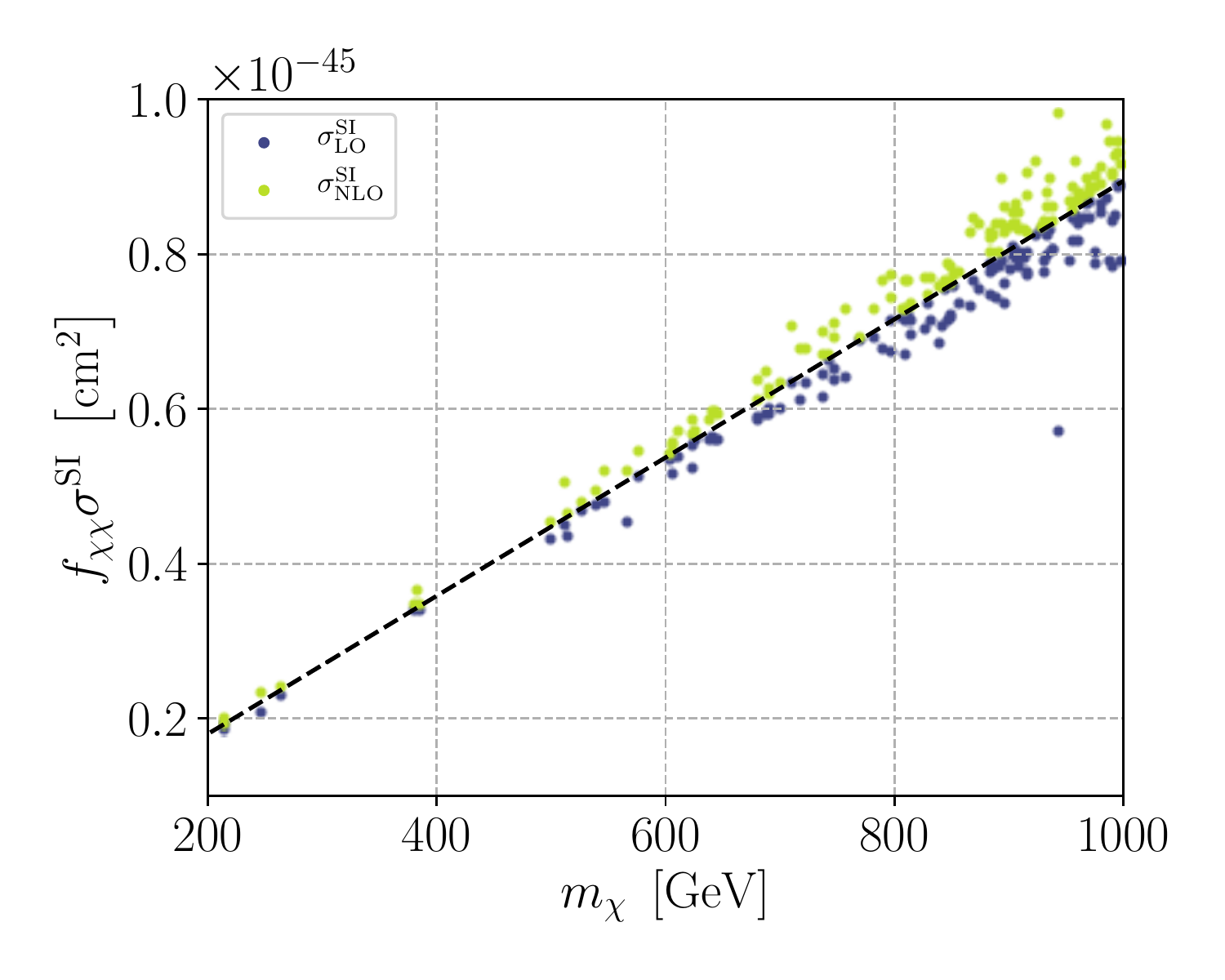}
    \caption{The SI cross section including the correction factor
      $f_{\chi\chi}$ at LO (blue) and NLO (green) compared to the
      Xenon limit (blue-dashed) versus the DM mass
      $\mX$.}
    \label{fig:result5}
\end{figure}

In order to understand the difference between LO and NLO at the phenomenological level we
present in~\cref{fig:result5} the SI cross section including the correction factor
$f_{\chi\chi}$ at LO (blue) and NLO (green) compared to the Xenon limit (blue-dashed) versus the DM mass $\mX$.  
In this plot we show all parameter points where the LO cross
section does not exceed the Xenon limit but the NLO result does. Clearly, there is
a sizeable number of parameter points where compatibility with the experimental constraints at NLO would no longer hold. 
Therefore, NLO corrections need to be accounted for in order to make reliable predictions about the viable parameter space of the VDM model.
%
It can also be that for some parameter points for which the LO cross
section is much smaller than the Xenon limit, the NLO cross section is of the order of the Xenon limit. In this case, although LO results might suggest that the Xenon experiment is
not sensitive to the model, this is no longer true when NLO corrections are taken into account. These results confirm the
importance of the NLO corrections when interpreting the data.

\section{Conclusions}\label{sec:conclusion}

This paper is an update to a previous work~\cite{Glaus:2019itb} where
we have computed the NLO corrections to the SI direct detection cross section
for the scattering of the VDM particle off a nucleon. This minimal model is an extension
of the SM with a vector dark matter particle and a new scalar that mixes with the
SM Higgs. Relative to our previous work we have included the contribution
of the NLO corrections to the lower vertex, that is, the $qqh$ vertex. This was
possible after we have devised a way to treat the IR divergences that appear
in these corrections. 

The overall conclusions are the same but the results are somewhat more stable
with the $K$-factor for NLO corrections being slightly
smaller.  There is clear hierarchy
in the significance of the NLO corrections where the leading role belongs
to both vertex corrections followed by mediator and finally by the box corrections.    
The interference effects between the two scalar particles, relevant for degenerate mass
values, were again found to be large and require further investigations beyond the 
scope of this paper. Outside this region, the perturbative series is
well-behaved. 

From the phenomenological point of view the overall conclusions are
again the same.
The NLO corrections can increase the 
LO results to values where the Xenon experiment becomes sensitive to
the model, or to values where the model is even excluded due to cross sections
above the Xenon limit. In case of suppression, parameter points that might be rejected at LO may render 
the model viable when NLO corrections are included.

\subsubsection*{Acknowledgments}
We are thankful to M.~Gabelmann, M.~Krause and
M.~Spira for fruitful and clarifying discussions. We are grateful to
D.~Azevedo for providing us with the data samples. R.S. is  
supported by the Portuguese Foundation for Science and Technology (FCT), 
Contracts UIDB/00618/2020, UIDP/00618/2020, PTDC/FIS-PAR/31000/2017
and CERN/FISPAR/0002/2017, and by the HARMONIA project, contract
UMO-2015/18/M/S. The work of MM is supported by the
  BMBF-Project 05H18VKCC1, project number 05H2018.
\vspace*{0.5cm}
\appendix
\section{Nuclear Form Factors}\label{APP::NUCLEAR}
We here present the numerical values for the nuclear form factors
defined in \cref{HISANO::MATRIXELEMENT}. The values of the form
factors for light quarks are taken from {\tt
  micrOmegas}\cite{Belanger:2018mqt} 
\begin{subequations}    
    \begin{align}
        & f^p_{T_u} = 0.01513\,,\quad f^p_{T_d} = 0.0.0191\,,\quad f^p_{T_s} = 0.0447\,, \\
        & f^n_{T_u} = 0.0110\,,\quad f^n_{T_d} = 0.0273\,,\quad f^n_{T_s} = 0.0447\,, 
    \end{align}
\end{subequations}
which can be related to the gluon form factors as
\begin{align}
    f^p_{T_G} = 1-\sum_{q=u,d,s} f_{T_q}^p\,,\qquad f^n_{T_G} = 1-\sum_{q=u,d,s} f_{T_q}^n\,.
\end{align}
The needed second momenta in \cref{HISANO::MATRIXELEMENT} are defined
at the scale $\mu=m_Z$ by using the {\tt CTEQ} parton distribution
functions \cite{Pumplin:2002vw},
\begin{subequations}    
    \begin{align}
        u^p(2) = 0.22\,,\qquad & \bar u^p(2) = 0.034\,, \\
        d^p(2) = 0.11\,,\qquad & \bar d^p(2) = 0.036\,, \\
        s^p(2) = 0.026\,,\qquad & \bar s^p(2) = 0.026\,, \\
        c^p(2) = 0.019\,,\qquad & \bar c^p(2) = 0.019\,, \\
        b^p(2) = 0.012\,,\qquad & \bar b^p(2) = 0.012\,, 
    \end{align}
\end{subequations}
where the respective second momenta for the neutron can be obtained by
interchanging up- and down-quark values.


\end{document}